\documentclass{jpp}
\usepackage{graphicx}

\usepackage[utf8]{inputenc}
\usepackage[T1]{fontenc}
\usepackage{amsmath}
\usepackage{float}
\usepackage[dvipsnames]{xcolor}
\usepackage[pdfauthor={A. Wiedman, S. Buller, M. Landreman},
            pdftitle={Coil Optimization for Quasi-helically Symmetric Stellarator Configurations},
            pdfsubject={coil optimization},
            pdfkeywords={coil optimization nuclear fusion stellarator}]{hyperref}

\renewcommand{\vec}[1]{\boldsymbol{#1}}

\shorttitle{Coil Optimization for Quasi-helically Symmetric Stellarator Configurations}
\shortauthor{A.~Wiedman, S.~Buller, M.~Landreman}

\title{Coil Optimization for Quasi-helically Symmetric Stellarator Configurations}

\author{A.~Wiedman\aff{1}, S.~Buller\aff{1}, M.~Landreman\aff{1}}

\affiliation{\aff{1}Institute for Research in Electronics and Applied Physics, University of Maryland, College Park, MD 20742, USA}

\begin{document}

\maketitle

\begin{abstract}
Filament-based coil optimizations are performed for several quasihelical stellarator configurations, beginning with the one from [M.~Landreman and E.~Paul, Phys.~Rev.~Lett.~\textbf{128}, 035001, 2022], demonstrating that precise quasihelical symmetry can be achieved with realistic coils.
Several constraints are placed on the shape and spacing of the coils, such as low curvature and sufficient plasma-coil distance for neutron shielding. The coils resulting from this optimization have a maximum curvature 0.8 times that of the coils of the Helically Symmetric eXperiment (HSX) and a mean squared curvature 0.4 times that of the HSX coils when scaled to the same plasma minor radius. When scaled up to reactor size and magnetic field strength, no fast particle losses were found in the free-boundary configuration when simulating 5000 alpha particles launched at $3.5\,\mathrm{MeV}$ on the flux surface with normalized toroidal flux of $s=0.5$.
An analysis of the tolerance of the coils to manufacturing errors is performed using a Gaussian process model, and the coils are found to maintain low particle losses for smooth, large-scale errors up to amplitudes of about $0.15\,\mathrm{m}$.
Another coil optimization is performed for the Landreman-Paul configuration with the additional constraint that the coils are purely planar. Visual inspection of the Poincar\'{e} plot of the resulting magnetic field-lines reveal that the planar modular coils alone do a poor job of reproducing the target equilibrium.
Additional non-planar coil optimizations are performed for the quasihelical configuration with $5\%$ volume-averaged plasma beta from [M.~Landreman, S.~Buller, and M.~Drevlak, Physics~of~Plasma.~\textbf{29} (8), 082501, 2022], and a similar configuration also optimized to satisfy the Mercier criterion. 
The finite beta configurations had larger fast-particle losses, with the free-boundary Mercier-optimized configuration performing the worst, losing about $5.5\%$ of alpha particles launched at $s=0.5$.

\end{abstract}

\section{Introduction}
Unlike toroidal confinement concepts with an explicit direction of continuous symmetry, stellarators need to be optimized to ensure that collisionless drift-orbits are confined.
One way to achieve such confinement is to optimize the magnetic field to be quasisymmetric. In a quasisymmetric field, the magnetic field strength $B$ is constant with respect to a linear combination of the Boozer angles -- special coordinates in which the particle dynamics only explicitly depend on the magnitude of $B$ instead of the full magnetic vector field $\vec{B}$. As a result, the particle orbits behave as if the field had an exact symmetry \citep{boozer1983}.
Depending the direction of symmetry in $B$, the contours of $B$ will close either toroidally, helically, or poloidally, and the field is said to be quasi-axisymmetric (QA), quasihelically symmetric (QH) or quasipoloidally symmetric (QP). 
Apart from true axisymmetry, exact quasisymmetry is believed to be impossible to achieve in the entire toroidal volume \citep{garrenBoozer2}, but it is possible to approximate quasisymmetry to high accuracy \citep{nuhrenberg1988quasi}.

Recent developments in stellarator optimization have led to a variety of new magnetic field configurations with record low quasisymmetry error \citep{PhysRevLett.128.035001}.  However, for these improvements to be of practical relevance, it is necessary to produce the configurations with discrete electromagnetic coils. The goal of this paper is to demonstrate that it is indeed possible to find coils for several recent QH configurations.

In the conventional approach to stellarator design, a magnetic configuration is first optimized for confinement and other properties (stage I optimization), and a second optimization (stage II optimization) is used to find coils that reproduce this optimized configuration to sufficient accuracy \citep{merkel1987, stickler2002}. Coils can be optimized either through current-potential methods, where currents are optimized on a specified winding-surface \citep{merkel1987, regcoil1}, or using filamentary coils, where the optimization is done directly on parameters describing the coil shapes and currents \citep{zhu2018}.

The latter is the approach taken here. Specifically, we search for optimized coils for the precise quasihelical vacuum configuration found by \citet{PhysRevLett.128.035001}, a similar 4-field-period QH configuration with $5\%$ volume-averaged plasma $\beta$ \citep{landremanBullerDrevlak2022}, and a similar $5\%$ $\beta$ QH configuration also optimized for Mercier stability.
Here, $\beta$ is the ratio of plasma thermal energy density to magnetic energy density.

The coils presented here are among the first to achieve precise quasihelical symmetry. 
While coils for QH configurations have been designed and even built in the past \citep{HSX,bader_atal_2020}, these configurations had significantly higher quasisymmetry error. Previously, coils producing several precisely quasi-axisymmetric fields were presented in \citet{wechsungPreciseQA2022}.
Coils for quasihelical symmetry were recently presented briefly in \citet{Roberg_Clark_2023}, although the field produced by these coils was not analyzed in detail. 

Much of the difficulty in coil optimization lies in the trade-off between accurately reproducing the fields and coil complexity. Coil complexity can be a major factor in construction cost of a stellarator \citep{ncsx_postmortem2012, klinger2013461}.
Coils also require shielding from high plasma temperatures and neutron bombardment, so a minimum distance from the plasma surface must be maintained. There should also be gaps between coils large enough to allow for access to the plasma by diagnostics and for maintenance, and to accommodate strain from the strong magnetic forces at work. Coil optimization must account for these different competing objectives in order to for coils to be practical. 

A downside of doing the coil optimization separately from the optimization of the magnetic configuration is that different configurations with similar performance may be easier to reproduce accurately with simple coils. For this reason, coil optimization is sometimes incorporated into the optimization of the plasma shape \citep{pomphrey2001, jorge2023, giuliani2023}.
However, this comes with the downside of needing to optimize a more complicated objective function with more parameters. An alternative approach may be to instead penalize aspects of the magnetic configuration that lead to more complicated coils, such as scale-lengths in the magnetic field \citep{kappel2023prep}.

Small error tolerances are another driver of costs in stellarator construction. The National Compact Stellarator eXperiment (NCSX) was cancelled due to the escalating costs associated with the precision required for the coils \citep{ncsx_postmortem2012}. Hence, it is important to investigate the robustness of coils with respect to errors.

The rest of the paper is structured as follows: in  \autoref{sec:stell_config} we present the QH configurations in more detail and describe some of their properties. In \autoref{sec:tools_and_optimization} we outline the method of optimization for finding accurate and simple coils. In \autoref{sec:errortol} we introduce a measurement of error tolerance and compute it for our coils. In \autoref{sec:results} we show a specific set of coils found for the vacuum case and give several metrics of comparison to the original QH configuration. We then present coils for two finite-$\beta$ QH configurations and evaluate their performance.
Finally, we summarize our results in \autoref{sec:conclusion}.

Our coils and optimization scripts are available as supplementary materials in Zenodo at \url{https://doi.org/10.5281/zenodo.10211349}, see \citet{wiedman_2023_10211349}.


\section{Method}
\subsection{Stellarator Configurations}\label{sec:stell_config}
The three QH stellarator configurations we will consider are all stellarator symmetric, with $n_{\text{fp}} = 4$ field periods. The configurations all have a volume-averaged magnetic field strength $B$ of $5.86\,\mathrm{T}$ and a minor radius $a=1.7\,\mathrm{m}$, to match the ARIES-CS reactor design \citep{AriesCS}.

The precise QH configuration from \citet{landremanBullerDrevlak2022} has an aspect ratio $A = 8.0$ and mean rotational transform $\iota = 1.24$, is a vacuum field, and is optimized purely for low quasisymmetry error and the achieved aspect ratio, without taking magneto-hydrodynamic (MHD) stability into account.

The second configuration, from \citet{PhysRevLett.128.035001}, has $A=6.5$ and volume-averaged $\beta=5\%$ and was optimized to have self-consistent bootstrap current and a rotational transform profile $\iota$ that avoids crossing $\iota=1$, but is otherwise not optimized for MHD stability.

The third configuration was generated using the same optimization method as in \citep{landremanBullerDrevlak2022}, but with the addition of a Mercier stability term to the objective function, resulting in Mercier stability throughout the plasma.
It is otherwise similar to the second configuration.
This configuration has not appeared in a previous publication.


\subsection{Computational Tools and Optimization objective}\label{sec:tools_and_optimization}
In order to perform an optimization, coils must be represented mathematically. Using the SIMSOPT framework \citep{simsopt2021}, we describe coils as curves in space determined by Fourier series for each of the three Cartesian coordinates of the coils. We neglect the volume of the coils and instead represent them as current-carrying filaments. This is a reasonable approximation for coils far from the plasma \citep{McGreivy_2021}, but is insufficient to account for the magnetic forces on the coils, where information about the cross sections of the coils must be supplemented \citep{hurwitz2023efficient, landreman2023efficient}.

For our filamentary coils, each Cartesian component of the curve is represented as
\begin{equation}\label{eq:Fourier}
\begin{aligned}
X(\varphi) = \sum_{m=0}^{N} A_m\cos{m \varphi} + \sum_{m=1}^{N} B_m\sin{m \varphi}
\end{aligned}
\end{equation}
where $\varphi$ is an angle-like variable parameterizing the coil, and $N$ is the number of Fourier modes used in the representation.

We optimize the coils with the limited-memory BFGS algorithm from the \textit{scipy} library \citep{2020SciPy-NMeth}. SIMSOPT uses Python for front-end programming and C++ for back-end calculations, resulting in a fast and user friendly environment.

Optimization of the coils is performed by minimizing the objective function
\begin{equation}\label{eq:obj}
\begin{aligned}
F_{min} = \frac{1}{2} \int_s (\vec{B} \cdot \vec{\hat{n}})^2 \mathrm{d}S
+ \sum_{i=1}^{6} f_i
\end{aligned} 
\end{equation}
where the first term corresponds to the square of the flux of the magnetic field created by the coils through the target plasma surface. Here $\vec{B}$ is the magnetic field created by the coils and $\vec{\hat{n}}$ is the unit vector normal to the target surface.
The $f_i$ terms are related to various aspect of the coils, and are defined as follows.
\begin{equation}\label{eq:length}
\begin{aligned}
f_{1} = w_{\mathrm{length}} \sum_i\frac{1}{2}(L_{i} - L_{0})^2,
\end{aligned} 
\end{equation}
where $L_i$ is the length of each coil and $L_0$ is a target coil length.
\begin{equation}\label{eq:ccdist}
\begin{aligned}
f_{2} = w_{\mathrm{CCDist}} \sum_i{\sum_j{\max(0, d_{\mathrm{cc}, 0} - \min(d_{i,j}))^2}},
\end{aligned} 
\end{equation}
where $\min{d_{i,j}}$ is the closest distance between any point on coil $i$ and any point on coil $j$ with a threshold value of $d_{\mathrm{cc}, 0}$ below which the penalty applies. This penalizes coils that are too close to one another.
\begin{equation}\label{eq:csdist}
\begin{aligned}
f_{3} = w_{\mathrm{CSDist}} \sum_i\max(0, d_{\mathrm{cs}, 0} - \min{(d_{\mathrm{cs}, i})})^2,
\end{aligned} 
\end{equation}
where $\min{d_{\mathrm{cs}, i}}$ is the minimum distance between any point on coil $i$ and the boundary magnetic surface, and $d_{\mathrm{cs}, 0}$ is a threshold distance. This term penalizes coils that are placed within $d_{\mathrm{cs}, 0}$ of the plasma surface.
\begin{equation}\label{eq:kappa}
\begin{aligned}
f_{4} = w_{\mathrm{curvature}} \sum_i\frac{1}{2} \int_{\mathrm{c}_i} \max(\kappa_i - \kappa_0, 0)^2 ~dl,
\end{aligned} 
\end{equation}
where $\kappa_i$ is the curvature on each coil and $\kappa_0$ is the threshold curvature above which the penalty applies, penalizing coils with high maximum curvature. The integral is over $l$, the arclength along the coil.
\begin{equation}\label{eq:msc}
\begin{aligned}
f_{5} = w_{\mathrm{MSC}} \sum_i \max\left(0,\left(\frac{1}{L}\int_{\mathrm{c}_i}{\kappa_i(l)^2dl}\right)-\mathrm{MSC}_0\right)^2,
\end{aligned} 
\end{equation}
where $\kappa_i(l)$ is the curvature along the length of coil $i$ and $\mathrm{MSC}_0$ is the threshold value, discouraging the optimizer from considering coils with high mean-squared curvature (MSC).
\begin{equation}\label{eq:linkWeight}
\begin{aligned}
f_{6} = w_{\mathrm{link}} \max\left(0,\left[\sum_i \sum_j \mathrm{Lk}(\mathrm{c}_i, \mathrm{c}_j)\right]-0.1\right)^2
\end{aligned} 
\end{equation}
where we use Gauss' linking number integral formula,
\begin{equation}\label{eq:link}
    \mathrm{Lk}(\mathrm{c}_i,\mathrm{c}_j) = \frac{1}{4\pi} \oint_{c_i}\oint_{c_j}\frac{\vec{r}_i - \vec{r}_j}{|\vec{r}_i-\vec{r}_j|^3} (d\vec{r}_i \times d\vec{r}_j)
\end{equation}
where $c_i$ is the first curve, $c_j$ is the second curve, and $\vec{r}_i$ and $\vec{r}_j$ are the associated position vectors \citep{linkingNumber}. This prevents interlocking coils from being considered. Since the numerical implementation of the above integral will not be exactly zero, we use a quadratic penalty in $f_6$ to suppress small values. The threshold value $0.1$ is taken to be much larger than the numerical errors and much smaller than $1$. Even though the discretization error is small, typically below 0.01, the total value of the objective function typically reaches $10^{-6}$ or lower. Using the "round" python function has the same effect.
Suppressing the errors in the linking number thus drastically improves the performance of the optimization. 

We use a normalized form of the error in the calculated magnetic field
\begin{equation}\label{eq:bn}
\frac{\langle|\vec{B}_{\text{coils}} \cdot \vec{\hat{n}}_{\text{target}}|\rangle}{\langle|\vec{B}_{\text{target}}|\rangle},
\end{equation}
as our measurement of how successfully the coils reproduce the target surface, where $\langle A \rangle$ is the flux surface average of $A$.
Another measure of success is whether energetic particle trajectories remain confined.
To this end, the guiding-center tracing code SIMPLE \citep{simple1,simple2} is used to calculate the loss fraction of high energy $\alpha$ particles.

\subsection{Error Tolerance}\label{sec:errortol}
Error tolerance tests are performed through the use of the \texttt{CurvePerturbed} class in SIMSOPT. A set of coils has their Cartesian coordinates perturbed through a Gaussian process that depends on the length scale $\ell$, a measurement of the smoothness of the random function generated, and a parameter $\sigma$ that scales the standard deviation of the perturbations. Following \citet{Wechsung_2022}, $\ell$ is kept at a constant value of $0.4\pi$, while $\sigma$ varies between $0.01\,\mathrm{m}$ and $0.1\,\mathrm{m}$. The Gaussian process model for the perturbations is described by the radial basis function kernel 
\begin{equation}\label{eq:cov}
    \mathrm{Cov}(X(\varphi_1), X(\varphi_2)) = \sum_{j=-\infty}^\infty \sigma^2 \exp\left(\frac{-(\varphi_1-\varphi_2+2\pi j)^2}{2\ell^2}\right),
\end{equation}
where shifting by all integers $j$ times $2\pi$ makes it periodic with period $2\pi$. (Note that the actual coil parameterization in SIMSOPT goes from 0 to 1). The series can be truncated at a low number of integers due to the fall off of the exponential function.
This exact kernel was previously used by \citet{Wechsung_2022} to describe coil errors. At $\ell=0.4\pi$, $\sigma = 0.01\,\mathrm{m}$ and $\sigma = 0.1\,\mathrm{m}$ correspond to coil errors with with average amplitudes of about $3\,\mathrm{cm}$ and $30\,\mathrm{cm}$, respectively.

After the perturbations are applied to our coils, we recalculate the metrics used to evaluate the quality of the coils.


\section{Results} \label{sec:results}

\subsection{Landreman-Paul Precise QH} 
\label{sec:Landreman-Paulresults}

\begin{figure}
    \centering
    \includegraphics[scale = 0.15]{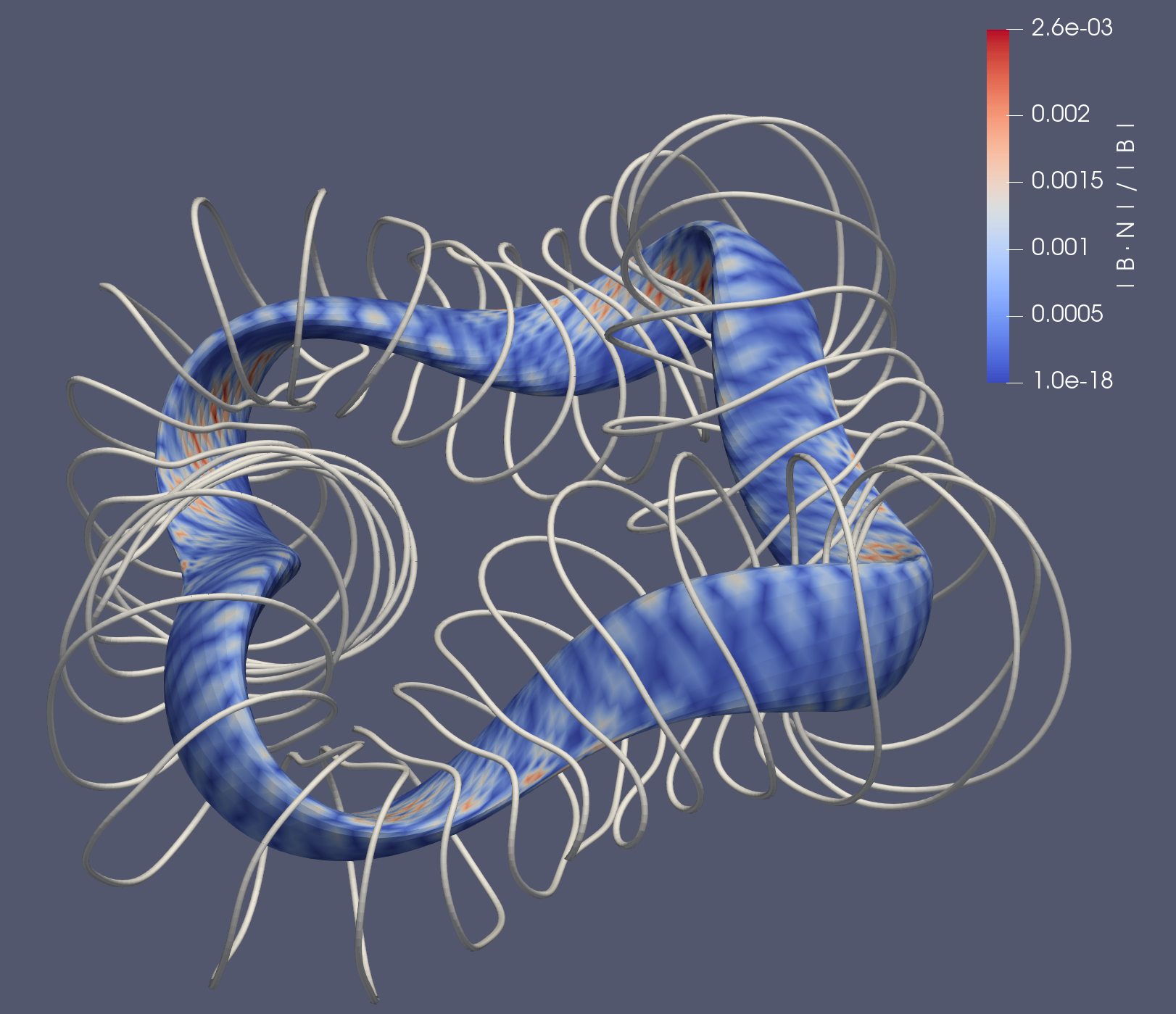}
    \caption{The target plasma surface of the Landreman-Paul precise QH configuration with the associated optimized coil set. The color scale shows the error in the $B$ field from the target surface as quantified by $|\vec{B} \cdot \vec{\hat{n}}|/B$.
    \label{fig:surf}}
\end{figure}

\begin{figure}
    \centering
    \includegraphics[width=\textwidth]{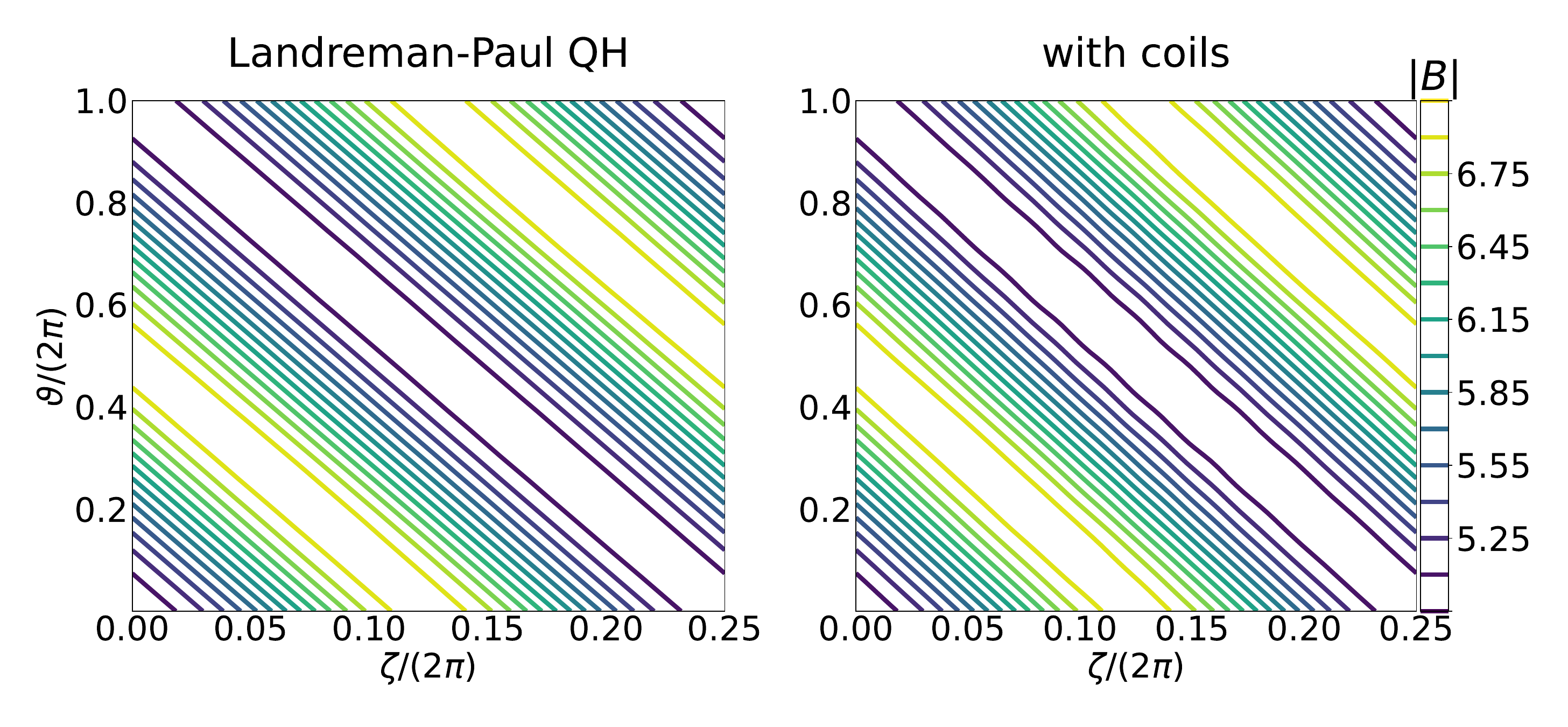}
    \caption{Magnetic field $B$ on the boundary in Boozer coordinates. Left plot: Landreman-Paul precise QH. Right plot: the $B$ produced by the coils presented in \autoref{sec:Landreman-Paulresults}.}
    \label{fig:countour}
\end{figure}

\begin{figure}
    \centering
    \includegraphics[width = \textwidth]{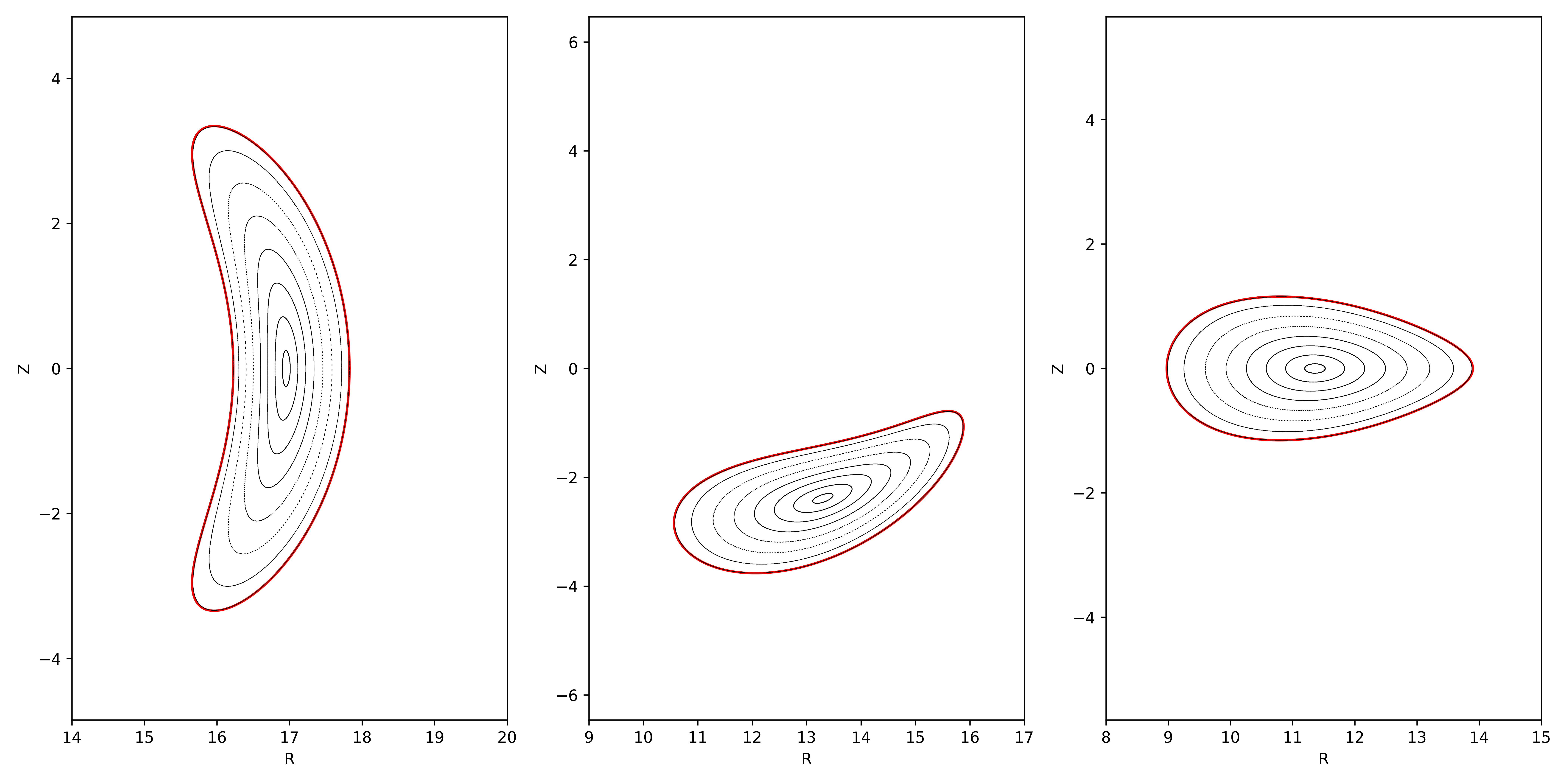}
    \caption{Poincar\'{e} plot of the flux surfaces created by the coils presented in \autoref{sec:Landreman-Paulresults}, at standard toroidal angle $\phi = 0, 1/4 $ period, $1/2$ period (from left to right). The red curve indicates the Landreman-Paul precise QH boundary targeted by the coils. The black lines depict the traced magnetic field lines within and on the boundary.}
    \label{fig:poincare}
\end{figure}

\begin{figure}
    \centering
    \includegraphics[width=0.45\textwidth]{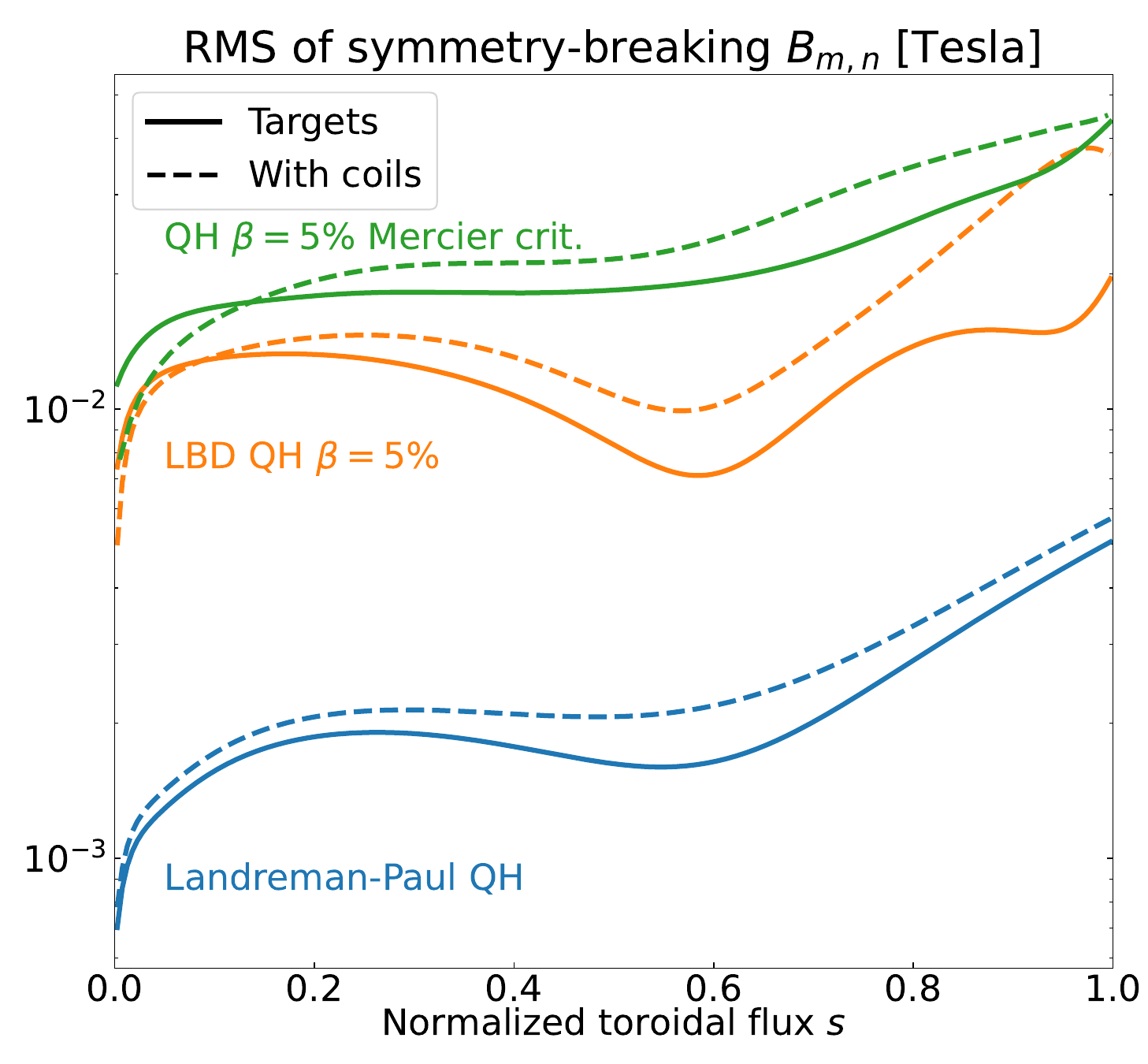}%
    \includegraphics[width=0.45\textwidth]{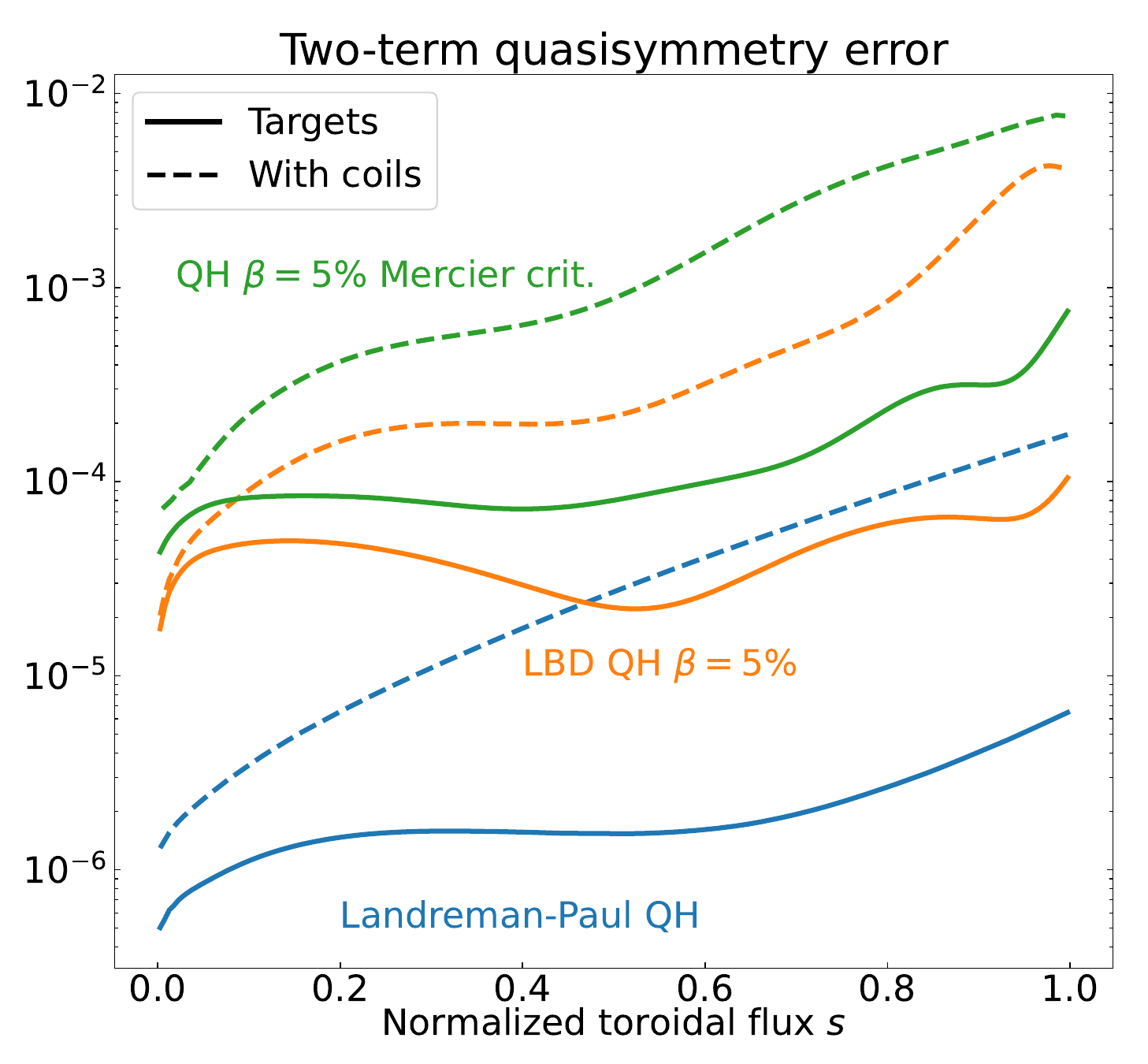}
    \caption{Quasisymmetry metrics for our configurations. (a) Root-mean-squared value of symmetry breaking Boozer harmonics $B_{m,n}$ against radius. (b) Two-term quasisymmetry error, as defined in \eqref{eq:2term}, against radius. Solid curves are for the target configurations, dashed curves for the configurations achieved by our coils. All configurations are scaled to match the volume-averaged $B$ and minor radius of ARIES-CS.
    \label{fig:qsmetrics}}
\end{figure}

\begin{table}
\begin{center}
    \begin{tabular}{r c c c c c c}
         & $L$ & CC Distance & CS Distance  & $\kappa$ & MSC &\\
        Weights & 0.0156 & 156 & 1560 & $1.56 \times 10^{-7}$ & $1.10 \times 10^{-8}$ &\\
        Target & 35.56\,m & 1.10\,m & 1.64\,m & $0.88\,\mathrm{m}^{-1}$ & $0.06\,\mathrm{m}^{-1}$ &\\
        Achieved Value & 35.56\,m & 1.09\,m &	1.62\,m & $0.67\,\mathrm{m}^{-1}$ & $0.07\,\mathrm{m}^{-1}$ &\\
        NCSX (scaled) & 36.60\,m & 0.79\,m & 1.00\,m & $1.68\,\mathrm{m}^{-1}$ & $0.40\,\mathrm{m}^{-1}$ & \\
        HSX (scaled) & 31.39\,m & 1.30\,m & 2.01\,m & $0.82\,\mathrm{m}^{-1}$ & $0.18\,\mathrm{m}^{-1}$ &\\
    \end{tabular}
    \caption{Weights placed on terms in objective function as described in equation \eqref{eq:obj} as well as the target values and the achieved values averaged across the set of coils, for the coil set described in \autoref{sec:Landreman-Paulresults}. The National Compact Stellarator eXperiment (NCSX) \citep{WILLIAMSON200571} and the Helically Symmetric eXperiment (HSX) \citep{HSX} metrics are shown as a comparison, scaled up to the plasma minor radius of ARIES-CS \citep{AriesCS}. Here $L$ is the average coil length (averaged across unique coils), $\kappa$ is the maximum curvature, and MSC is the mean squared curvature. The target of $\kappa$ being greater than the achieved value means the maximum curvature had no effect on the cost function at the end, however the minimizer was able to move through different values while searching for different minima, and the results were different than they would be for different target values.}
  \label{tab:weights}
\end{center}
\end{table}

We first optimized coils for the precise QH vacuum configuration presented in \autoref{sec:stell_config} using the method outlined in \autoref{sec:tools_and_optimization}, trying various weights for the objective terms $f_i$ in \eqref{eq:obj}. The most successful set of coils is shown in \autoref{fig:surf}. This set of coils was obtained using weights outlined in \autoref{tab:weights}. 

\begin{table}
\begin{center}
    \begin{tabular}{r c c c c}
         & $L$\,(m) & $\kappa$\,($\mathrm{m}^{-1}$) & MSC\,($\mathrm{m}^{-1}$) &\\
        Coil1 & 35.73 & 0.563 & 0.062 &\\
        Coil2 & 35.42 & 0.810 & 0.065 &\\
        Coil3 & 35.15 & 0.743 &	0.067 &\\
        Coil4 & 35.16 & 0.625 & 0.070 &\\
        Coil5 & 36.30 & 0.588 & 0.079 &\\
    \end{tabular}
    \caption{Length and curvature metrics of each unique coil in the coil set presented in \autoref{sec:Landreman-Paulresults}. There are 5 coils in the set, resulting in 40 total coils due to 4-field-period symmetry and stellarator symmetry. See \autoref{tab:weights} for definitions.}
  \label{tab:coils}
\end{center}
\end{table}

The set contains 5 coils per half period, 40 coils in total. This number of coils was chosen for the higher accuracy it was able to provide over 4 coil configurations, while still being able to maintain the desired coil to coil separation, which is much more difficult with a 6 coil configuration. Metrics for the five individual coils are presented in \autoref{tab:coils}. The coil lengths average to 35.56\,m. They maintain a distance of 1.09\,m between one another at minimum, and a minimum distance of 1.62\,m to the boundary surface. Each coil has a maximum curvature between 0.6\,$\mathrm{m}^{-1}$ and 0.8\,$\mathrm{m}^{-1}$, with coils having a range of MSC between $0.062\,\mathrm{m}^{-1}$ and $0.079\,\mathrm{m}^{-1}$, averaging at 0.069\,$\mathrm{m}^{-1}$. The target coil length, coil to coil distance, coil to surface distance, and curvature were all achieved or surpassed, while the mean squared curvature was slightly higher than the target. These values are all comparable to those of NCSX and HSX when scaled up to the same minor radius, with the curvature and MSC metrics of our coil set being lower.

The normalized flux surface average of the $B$ field from the coils normal to the target surface \eqref{eq:bn} is $6 \times 10^{-4}$ while the maximum $\vec{B}_{\text{coils}} \cdot \vec{\hat{n}}_{\text{target}}/|\vec{B}_{\text{target}}|$ is $3.1\times 10^{-3}$. \autoref{fig:surf} shows the error in $\vec{B} \cdot \vec{\hat{n}}$ on the surface. \autoref{fig:countour} and \autoref{fig:poincare} both give clear indication of the high degree of accuracy with which the coils recreate the target surface. There is some coil ripple present in \autoref{fig:countour} which is only present in the weaker section of the magnetic field, but the quasihelical symmetry is largely maintained by the coils.
The blue curves in \autoref{fig:qsmetrics}(a) show the root-mean-squared amplitude of the quasisymmetry breaking modes as a function of radius, for the target configuration and the configuration achieved with our coils. \autoref{fig:qsmetrics}(b) shows the corresponding curves for the two-term quasisymmetry error \citep{helanderSimakov2008}, defined as
\begin{equation}
f_{\text{QS}} =\left\langle \left(\frac{1}{B^3} [N - \iota M] \vec{B} \times \nabla B \cdot \nabla \psi - [MG + NI] \vec{B} \cdot \nabla B\right)^2\right\rangle, \label{eq:2term}
\end{equation}
where $(M,N)=(1,-4)$ corresponds to the helicity of the helical symmetry of $B$; $2\pi G/\mu_0$ and $2\pi I/\mu_0$ are the poloidal current outside and toroidal current inside the flux surface, respectively. Note that the factor $4$ in $N$ is due to $n_{\text{fp}} = 4$.

From comparing figure \autoref{fig:qsmetrics}(b) to \autoref{fig:qsmetrics}(a), we see that the two-term quasisymmetry errors exhibit much larger differences between the original stage I configurations and the configurations reproduced by our coils, compared to the differences in their RMS values. 
As shown by \citet{rodriguez_paul_bhattacharjee_2022}, the two-term form of quasi-symmetry can be written as a weighted version of the RMS value of symmetry-breaking Boozer modes. Specifically, for symmetry helicity $(M,N)=(1,-4)$, modes are weighted according to
\begin{equation}
  \left(\frac{n+4m}{\iota + 4}\right)^2.
\end{equation}
Thus, deviations from quasisymmetry with higher mode numbers, such as those created by modular coil ripple, are weighted more strongly in the two-term measure, which would explain the larger differences in \autoref{fig:qsmetrics}(b). Hence, the extent which the coils yield configurations that reproduce the quasisymmetry of the target configuration depends on which metric is used to quantify the degree of quasisymmetry. A more physically relevant performance metric is thus needed.

To evaluate the performance of our coils, we calculate how well the resulting configuration confine fast particles.
The guiding-center code SIMPLE is used to trace 5000 alpha particles with energies of $3.5\,\mathrm{MeV}$ for $0.2$ seconds, typical of the slowing-down time in a reactor.
Particles are launched at the flux surfaces with normalized toroidal flux $s = 0.3$, $s = 0.5$, and $s=0.7$, and are considered lost when they cross the $s=1.0$ surface, corresponding to the last closed flux surface. SIMPLE traces the collisionless guiding-center orbits, so the energy of the alpha particles remain constant.
For our coils, we find no fast-particle losses for particles launched at flux surfaces with normalized toroidal flux $s = 0.3$ and $s = 0.5$, and only $1.44\%$ losses when launched at $s=0.7$. For reference, the original precise QH configuration loses $1.40\%$ of particles launched at $s=0.7$,
indicating that the field is reproduced to sufficient accuracy.

\begin{figure}\centering
\includegraphics[width=0.5\textwidth]{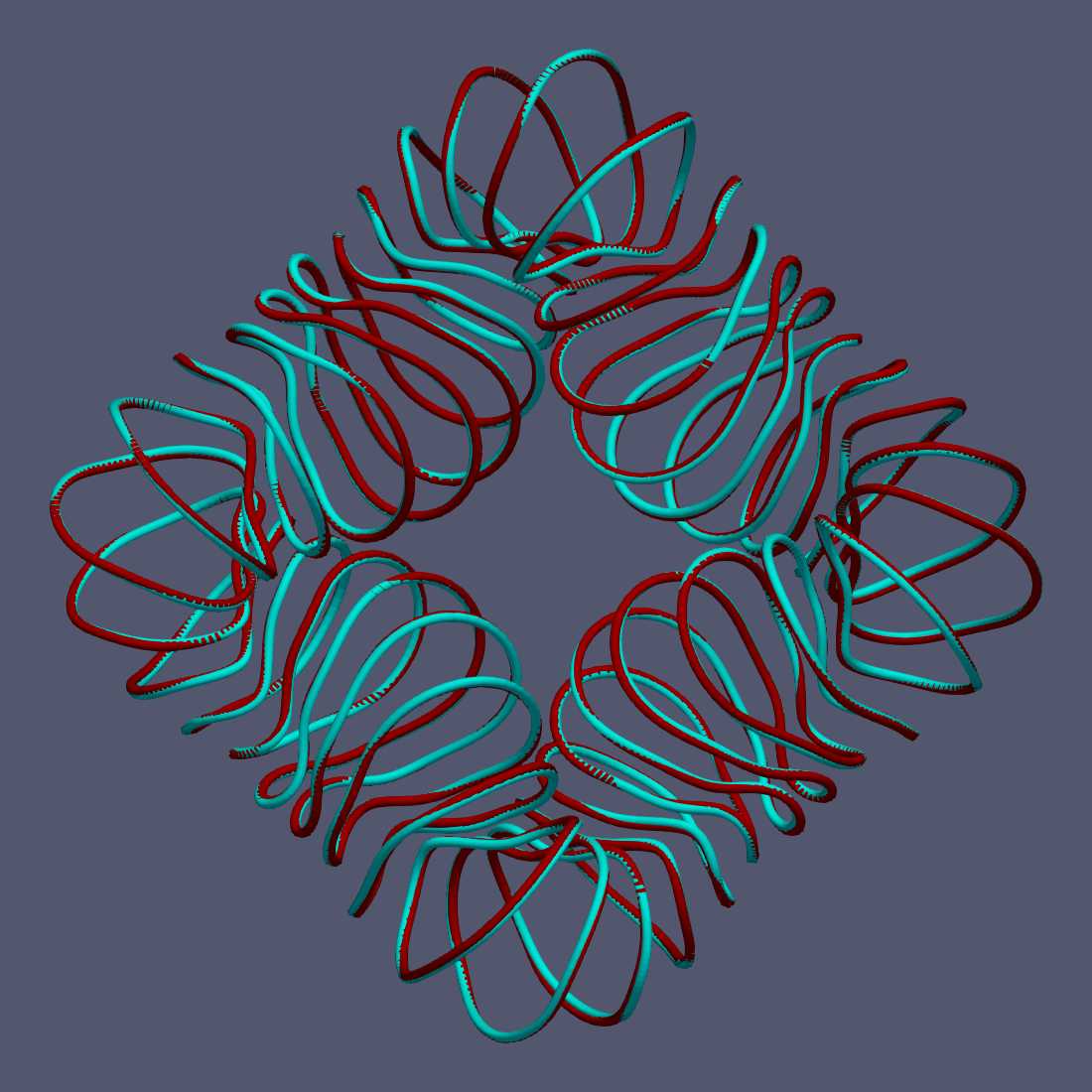}
\caption{\label{fig:oVp} The original coils (red) compared to a set of perturbed coils with $\sigma = 0.1\,\mathrm{m}$ (blue).}
\end{figure}

\begin{figure}
    \centering
    \includegraphics[width = \textwidth]{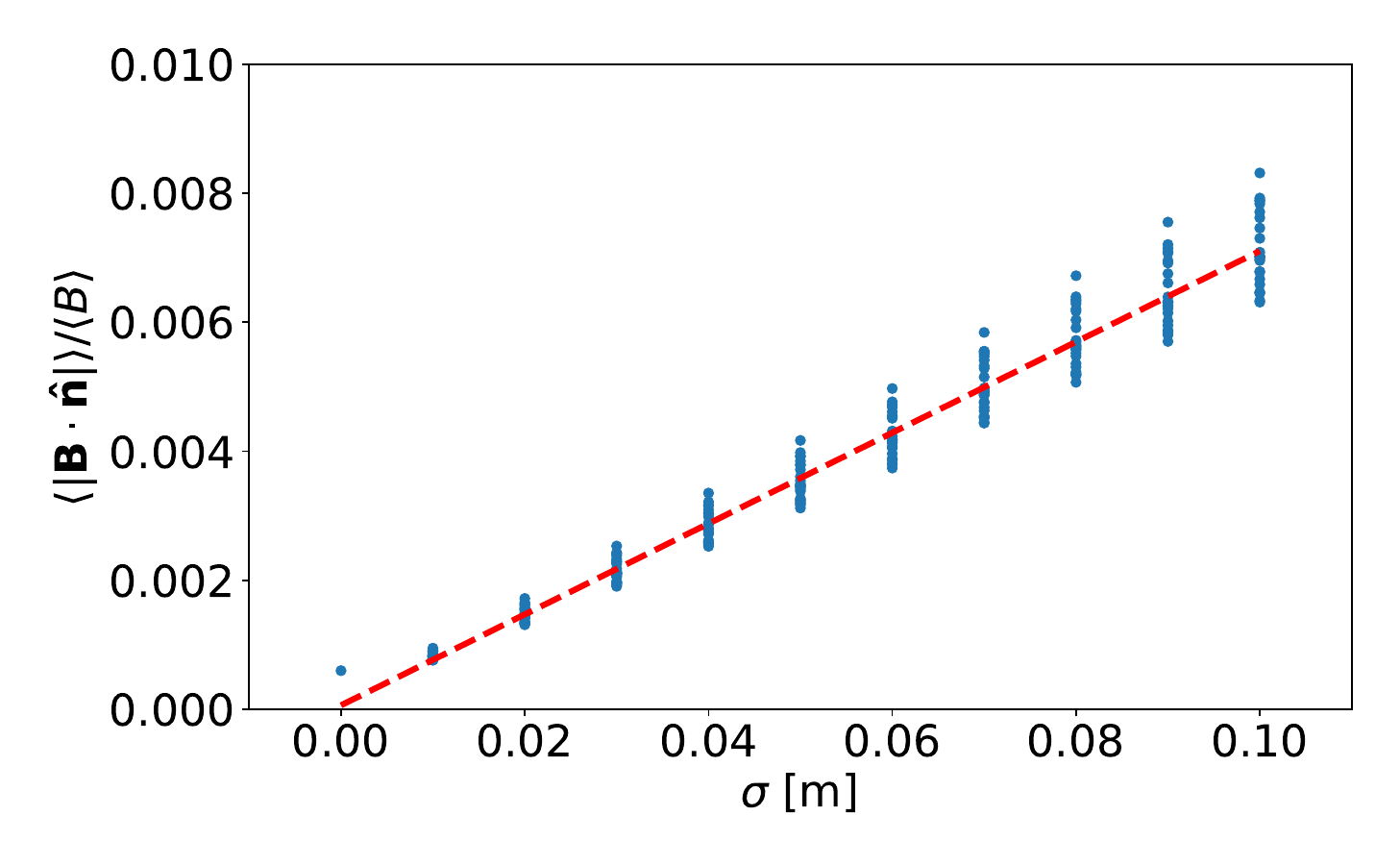}
    \caption{Scatter plot of the coil perturbation magnitude $\sigma$ versus the normalized flux surface average of $\vec{B}\cdot \vec{\hat{n}}$  (equation \ref{eq:bn}), with $\vec{B}$ calculated from perturbed coils targeting the Landreman-Paul precise QH configuration. The line indicates the least-squares regression for the discrete values of $\sigma$. }
    \label{fig:perturbed}
\end{figure}

\begin{figure}
    \centering
    \includegraphics[width = \textwidth]{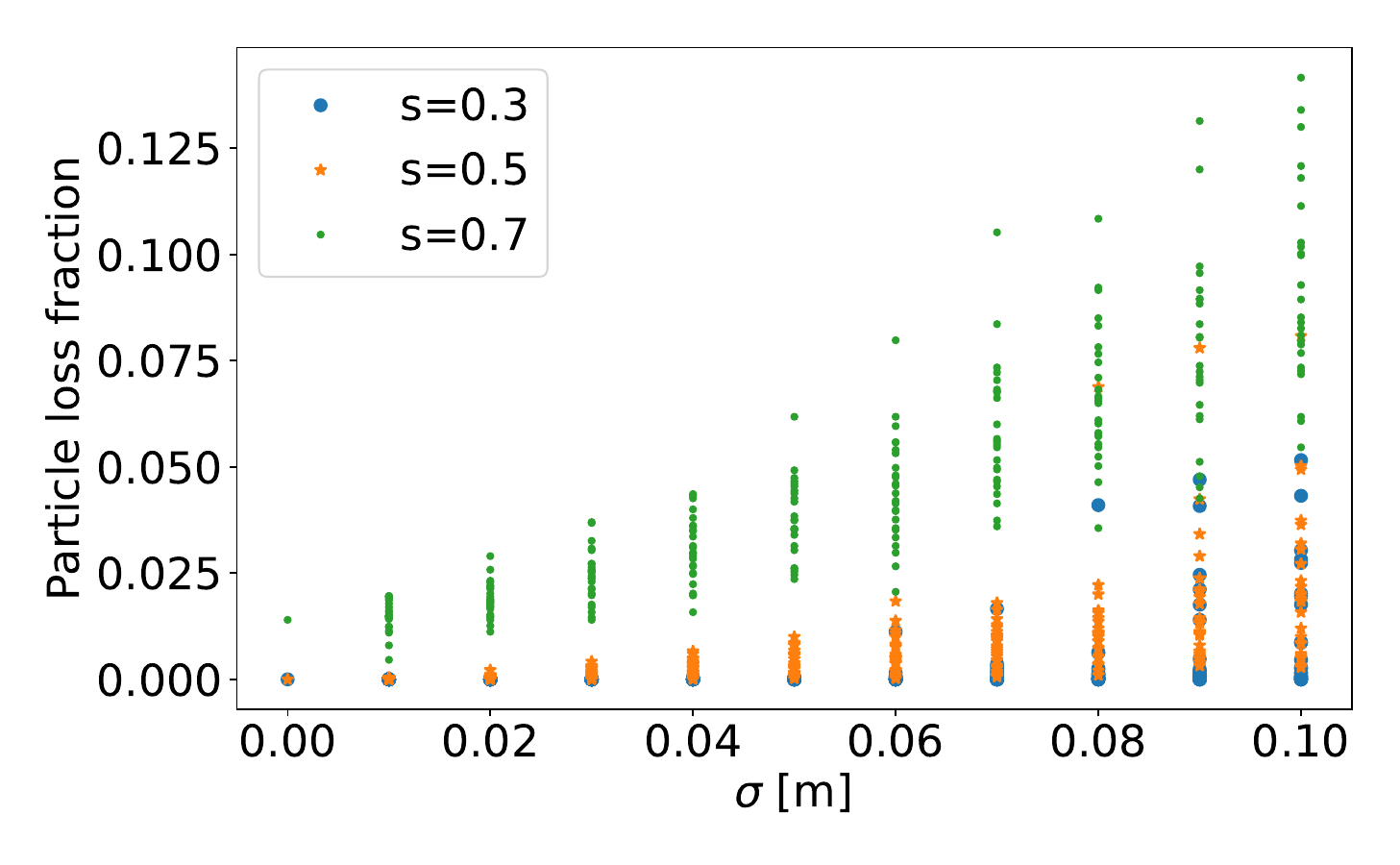}
    \caption{Fraction of high energy alpha particles lost when launched from different toroidal radii, for the coils perturbed from the set optimized for the Landreman-Paul precise QH in \autoref{fig:perturbed}. 
    }
    \label{fig:losses}
\end{figure}

To assess the error tolerance of the coil set, we generated 25 sets of coil perturbations, using the Gaussian process model described in \autoref{sec:errortol}. The generated perturbations were scaled to different amplitudes by using 10 different values for $\sigma$. The perturbations were then added to our optimal set of coils, resulting in a total of 250 perturbed coil sets. \autoref{fig:oVp} shows a perturbed coil set, with the unperturbed coil set included for reference.
The resulting flux surface averages of the $B$ field from the coils normal to the target surface are shown in \autoref{fig:perturbed}, while the corresponding fast-particle losses are shown in \autoref{fig:losses}.
At $\sigma = 0.1\,\mathrm{m}$, there is a perceptible difference in the shape of the coils, and between $0\% $  and $5\%$ of particles launched from a toroidal radius of $s=0.3$ are lost. 
Thus, there is considerable variation in particle confinement for a given $\sigma$.

\begin{figure}
    \centering
    \includegraphics[width =\textwidth]{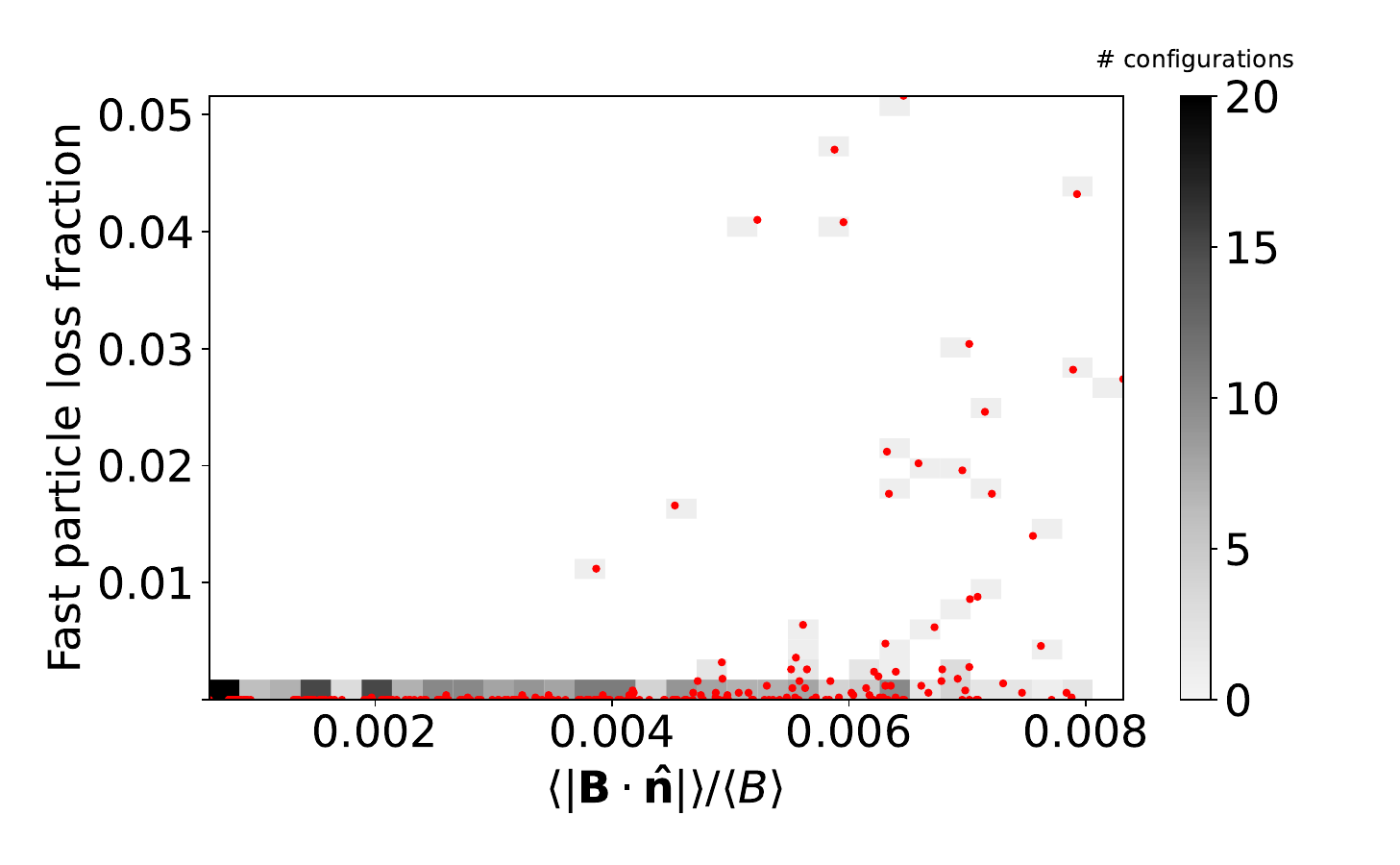}
    \caption{Scatter plot and 2D-histogram of the loss fraction of high energy alpha particles launched at $s=0.3$, against the flux surface average $\vec{B} \cdot \vec{\hat{n}}$, for the coils perturbed from the set optimized for the Landreman-Paul precise QH configuration.}
    \label{fig:bdotnVloss}
\end{figure}

\begin{figure}
    \centering
    \includegraphics[width =\textwidth]{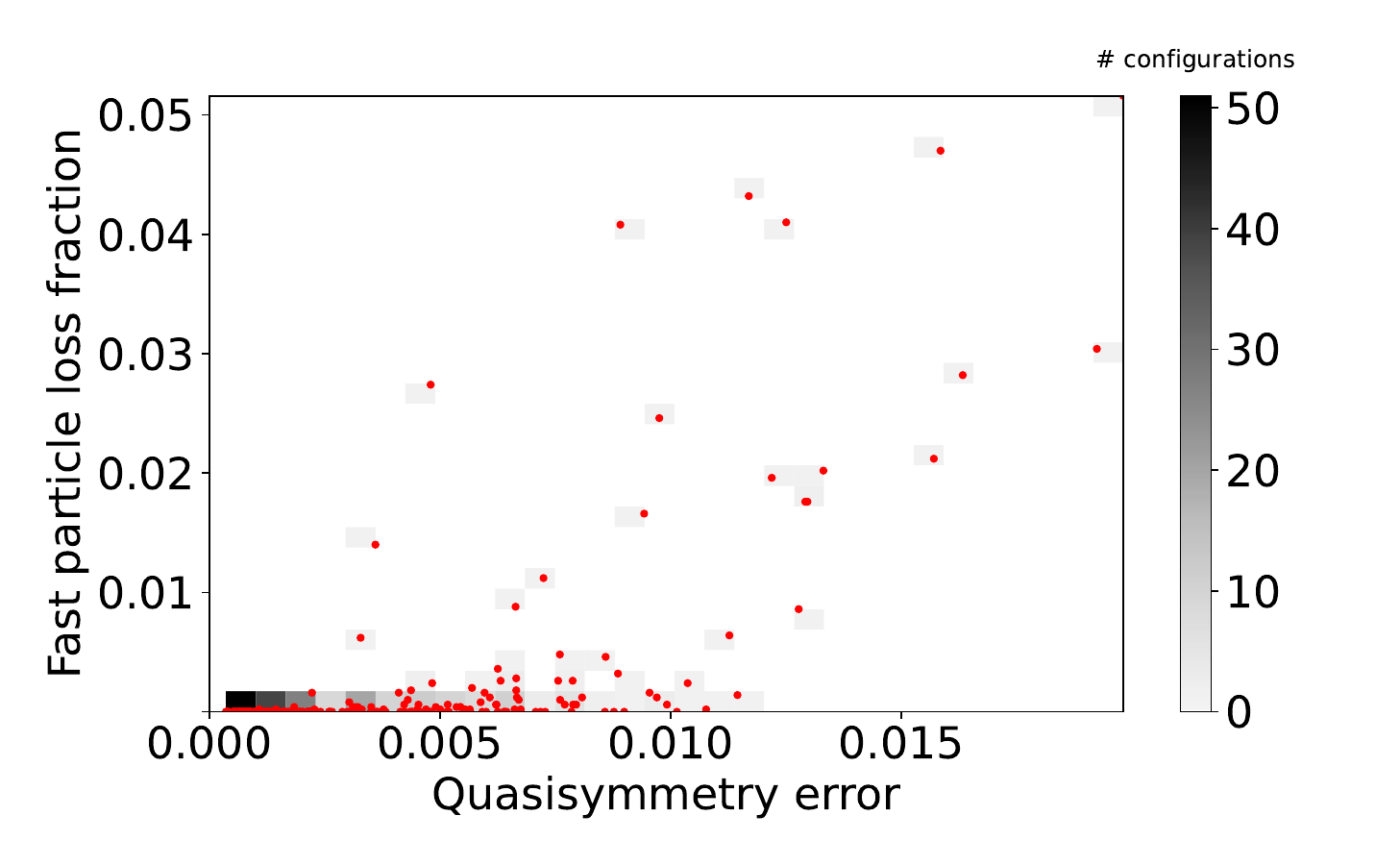}
    \caption{Scatter plot and 2D-histogram of the loss fraction of high energy alpha particles launched at  $s=0.3$, against different values of quasisymmetry error for the coils perturbed from the set optimized for the Landreman-Paul precise QH configuration.}
    \label{fig:qsVloss}
\end{figure}

To better understand the large spread in fast-particle losses, we plot the losses for particles launched from $s=0.3$ against both the error in the target field, the flux surface average (fig.~\ref{fig:bdotnVloss}) and the quasisymmetry error (fig.~\ref{fig:qsVloss}).
In \autoref{fig:qsVloss}, we sum the quasisymmetry error defined \eqref{eq:2term} for each of the flux surfaces $s = \{0, 0.1, \dots, 1.0\}$, and thus obtain a scalar measure of the quasisymmetry error in the entire plasma volume from axis to edge.

In both \autoref{fig:bdotnVloss} and \autoref{fig:qsVloss}, poor confinement is generally only found above certain thresholds, for $\langle |\vec{B}\cdot\vec{\hat{n}}|\rangle/\langle B\rangle$ above $0.5\%$ and quasisymmetry error above $0.003$, but configurations with good confinement are also found well above these values. As pointed out by \citet{rodriguez_paul_bhattacharjee_2022}, decreasing the two-term quasisymmetry error does not necessarily lead to a decrease in performance (or even a decrease in other measures of quasisymmetry error). This is consistent with our findings here.

\subsection{Planar Coils}
In order to explore how well lower-complexity coils can reproduce the target magnetic field, we also consider whether the same
plasma configuration can be produced with modular coils that are  planar. 

Each coil curve is taken to lie on a plane, with the distance of the coil from a central point described using a Fourier series. The rotation of the plane is described with a quaternion, and the position of the central point is specified using Cartesian coordinates. Quaternion rotation is used to prevent gimbal locking during optimization. The quaternion has 4 elements $[q,i,j,k]$ which represent a rotation about some vector $\vec{u}$ through the central point:
\begin{equation}\label{quaternion}
    \begin{aligned}[]
    [q, i, j, k] = [\cos{\frac{\theta}{2}}, \sin{\frac{\theta}{2}}u_x,\sin{\frac{\theta}{2}}u_y,\sin{\frac{\theta}{2}}u_z]
    \end{aligned}
\end{equation}
where $\theta$ is the rotation about $\vec{u}$. We restrict $\vec{u}$ to be a unit vector; if it is not, the rotation would change the length of the curve.

The position vector of a point along the coil before rotation is determined by
\begin{equation}\label{posVecX0}
    \begin{aligned}[]
    X_0(\varphi) = \sum_{m=0}^{N}(r_{c,m} \cos(m\varphi) \cos(\varphi)) + \sum_{m=1}^{N}(r_{s,m} \sin(m\varphi) \cos(\varphi))
    \end{aligned}
\end{equation}
\begin{equation}\label{posVecY0}
    \begin{aligned}[]
    Y_0(\varphi) = \sum_{m=0}^{N}(r_{c,m} \cos(m\varphi) \sin(\varphi)) + \sum_{m=1}^{N}(r_{s,m} \sin(m\varphi) \sin(\varphi))
    \end{aligned}
\end{equation}
\begin{equation}\label{posVecZ0}
    \begin{aligned}[]
    Z_0(\varphi) = 0
    \end{aligned}
\end{equation}
The quaternion rotation is then performed
\begin{equation}\label{posVecRot}
    \begin{aligned}[]
        \begin{bmatrix}
        X(\varphi) \\
        Y(\varphi) \\
        Z(\varphi)
        \end{bmatrix}
        =
        \begin{bmatrix}
        1 - 2(j^2+k^2) & 2(ij-kq) & 2(ik+jq)\\
        2(ij+kq) & 1-(i^2+k^2) & 2(jk-iq) \\
        2(ij-kq) & 2(jk+iq) & 1 - 2(i^2+j^2)
        \end{bmatrix}
        \begin{bmatrix}
        X_0(\varphi) \\
        Y_0(\varphi) \\
        Z_0(\varphi)
        \end{bmatrix}
        +
        \begin{bmatrix}
        x_{\mathrm{center}} \\
        y_{\mathrm{center}} \\
        z_{\mathrm{center}}
        \end{bmatrix}
    \end{aligned}
\end{equation}
where the matrices in \eqref{posVecRot} represent the position vector after rotation, the quaternion based rotation matrix as described in \citet{shepperd1978quaternion}, the position vector before rotation defined by \eqref{posVecX0}, \eqref{posVecY0}, \eqref{posVecZ0}, and a translation of the center.

To attempt to find an adequately low-error coil set, we performed 115 optimizations while varying the weights and threshold values as described in \eqref{eq:obj}. The weights and threshold values were varied around the previously found optimum values, so that we only consider planar coils of similar complexity to the non-planar coils.

\begin{figure}
    \centering
    \includegraphics[width =0.5\textwidth]{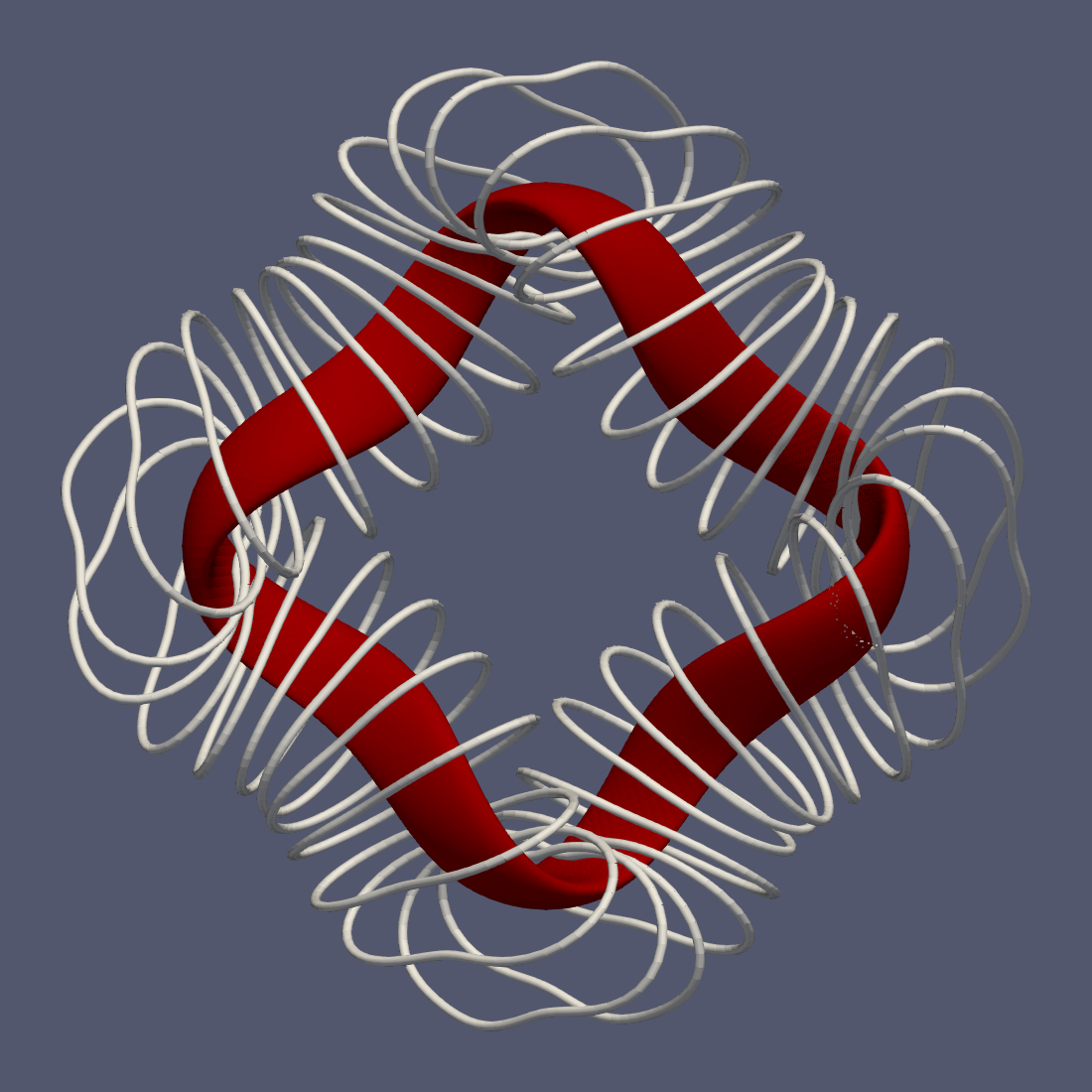}
    \caption{Planar coils optimized for the Landreman-Paul precise QH, with the target surface (red).}
    \label{fig:planarCoils}
\end{figure}
\begin{figure}
    \centering
    \includegraphics[width =\textwidth]{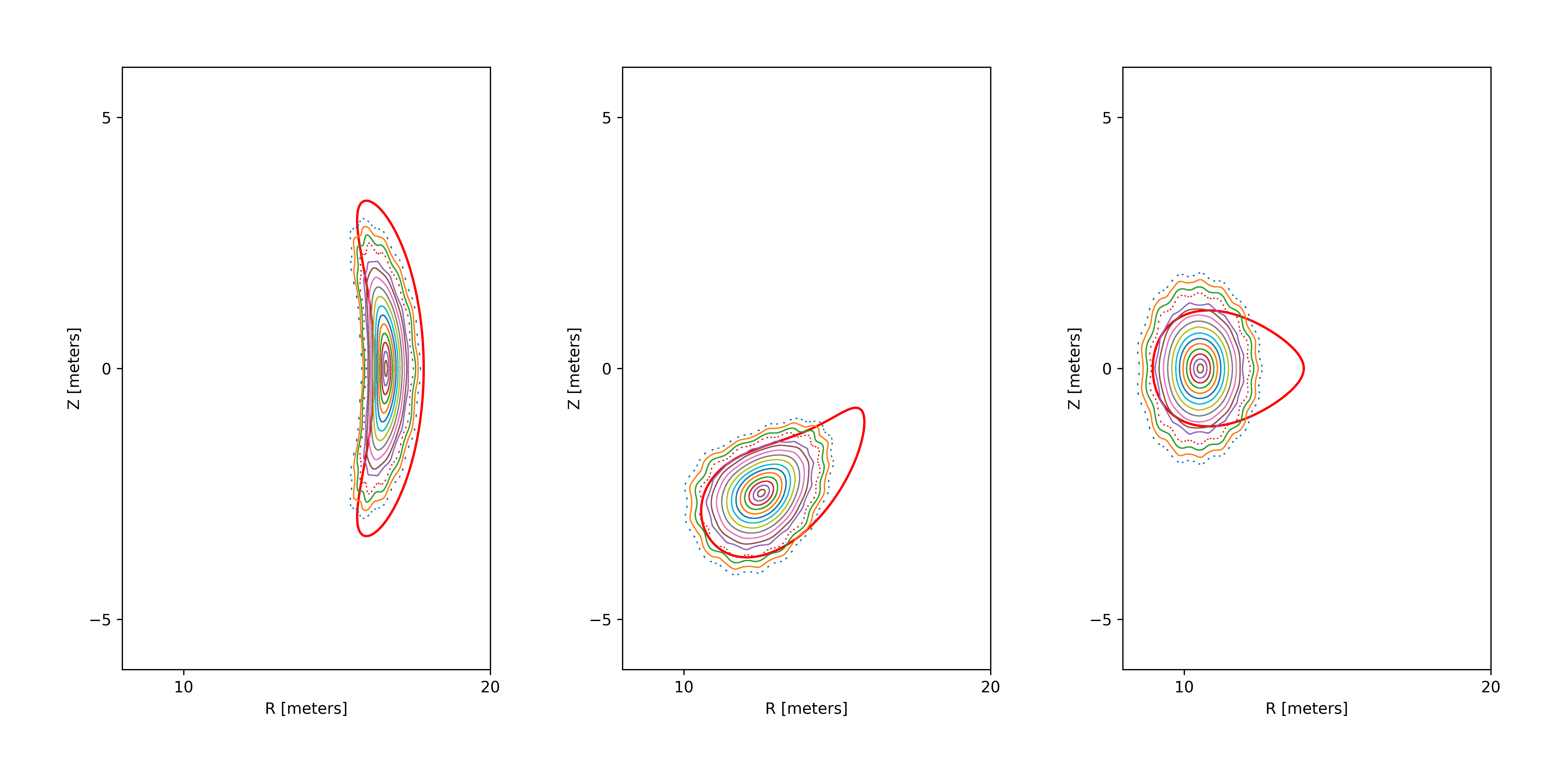}
    \caption{Poincar\'{e} plot of the flux surfaces created by the planar coils optimized for the Landreman-Paul precise QH. Cross sections are taken at standard toroidal angle $\phi = 0, 1/4 $ period, $1/2$ period (from left to right).
    The solid red curves indicate the target plasma boundary.}
    \label{fig:planarPoincare}
\end{figure}

The lowest normalized flux surface average of $|\vec{B} \cdot \vec{\hat{n}}|$ (equation \ref{eq:bn}) thus obtained was 0.0382. The average length of the resulting coils is 40.2\,m, the maximum curvature and MSC were $0.397\,\mathrm{m}^{-1}$ and $0.577\,\mathrm{m}^{-1}$ respectively, while the distance between coils and the distance from the coils to the surface were kept above 1.39\,m and 0.96\,m respectively. The coils are shown in \autoref{fig:planarCoils}.
A Poincar\'{e} plot of the resulting magnetic field is shown in \autoref{fig:planarPoincare}. As seen from the Poincar\'{e} plot, the coils do a poor job of approximating the target surface.
While producing this QH configuration with only planar modular coils does not appear possible, planar modular coils may be feasible for other plasma configurations, or perhaps if used along with windowpane coils.
These possibilities are left for future work.



\subsection{Landreman-Buller-Drevlak $5\%$ volume-averaged $\beta$ QH\label{sec:LBDQH5}}
\begin{figure}
    \centering
    \includegraphics[width = 1.0\textwidth]{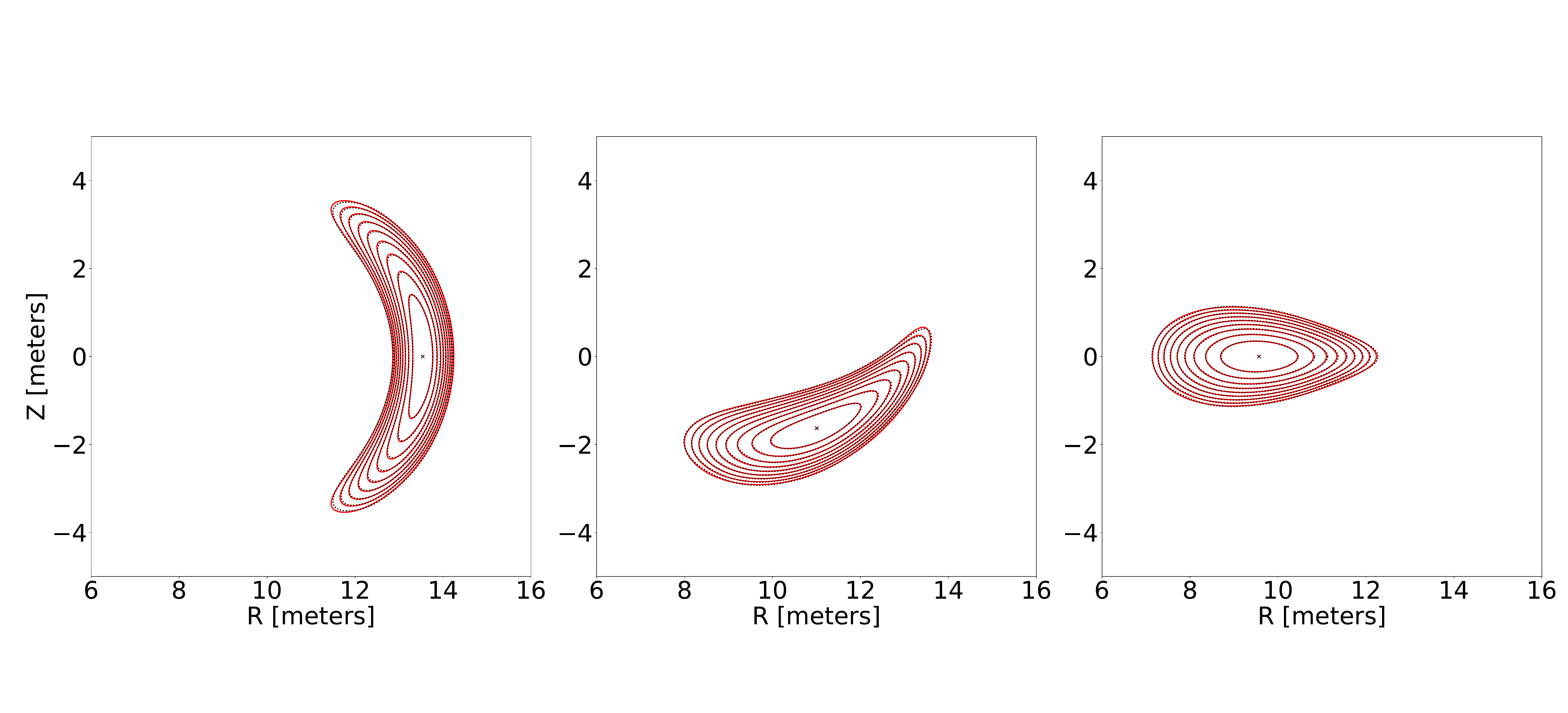}
    \caption{Flux surface comparison of the Landreman-Buller-Drevlak QH $5\%$ volume-averaged $\beta$ configuration (red solid), and the surfaces produced by the coils (black dotted).}
    \label{fig:fluxsurfaces2}
  \end{figure}

  \begin{figure}
    \centering
    \includegraphics[width=\textwidth]{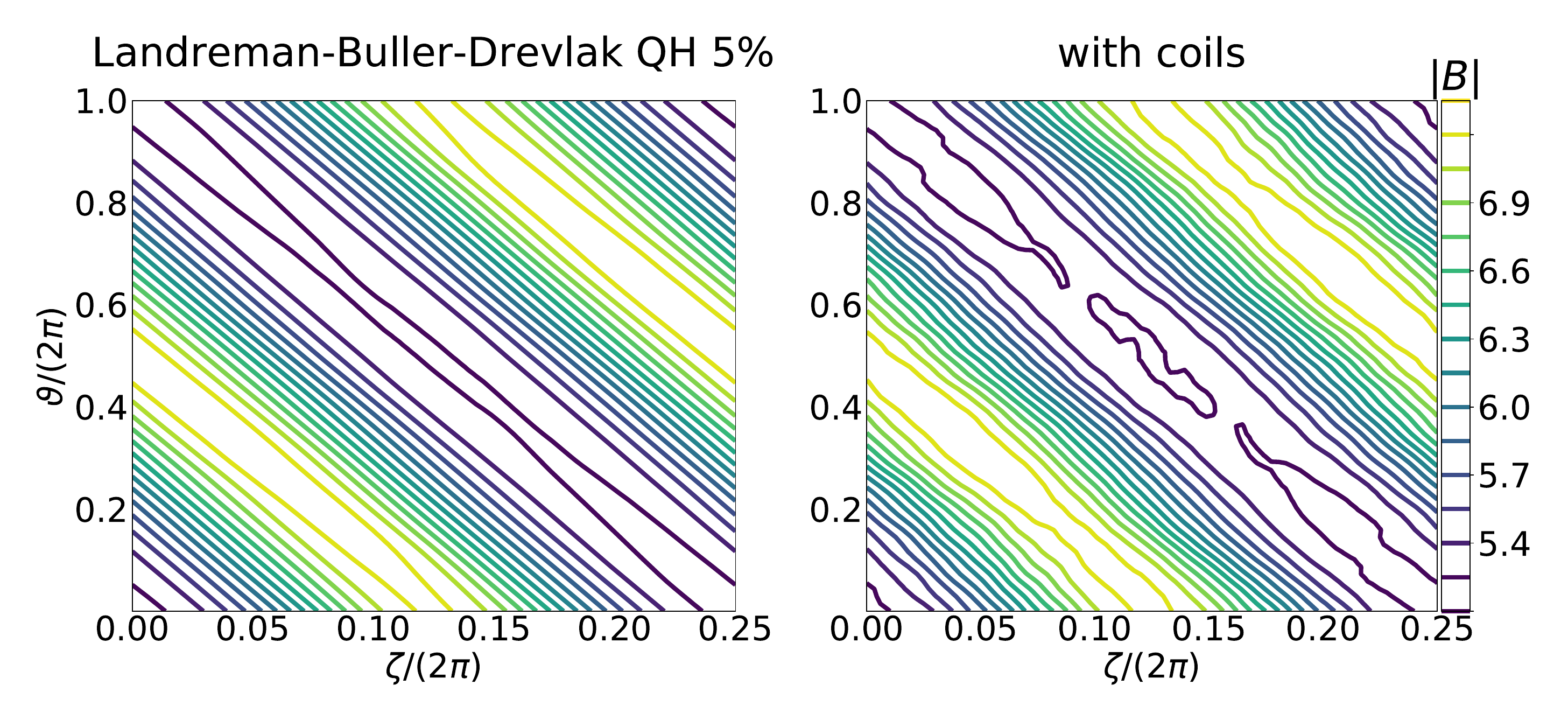}
    \caption{\label{fig:countourLBDQH5} Magnetic field $B$ on the plasma boundary in Boozer coordinates. Left plot:  Landreman-Buller-Drevlak QH with $5\%$ volume-averaged $\beta$. Right plot: $B$ produced by the coils presented in \autoref{sec:LBDQH5}.}
\end{figure}

\begin{figure}
    \centering
    \includegraphics[width = 0.5\textwidth]{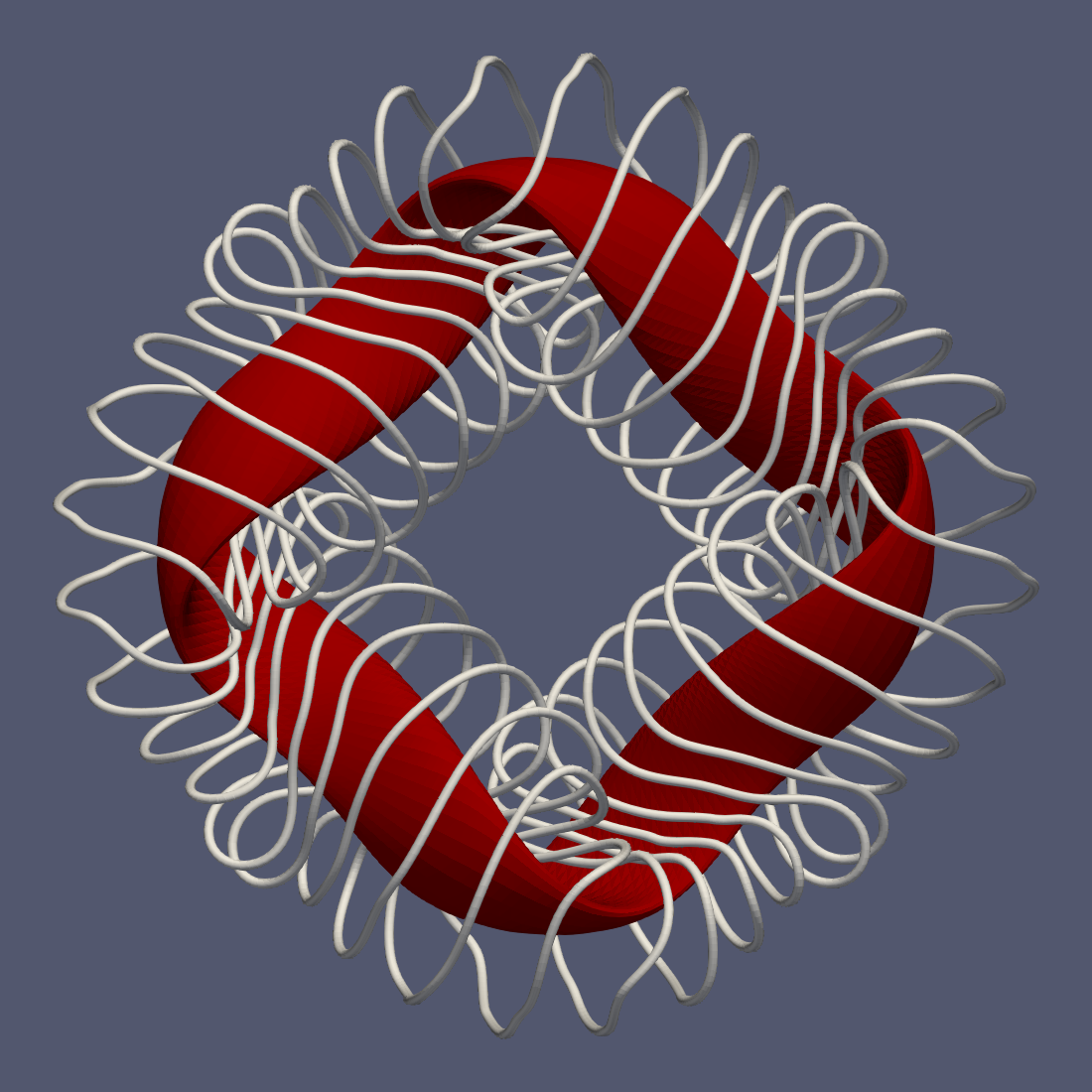}
    \caption{Optimized coil set for the Landreman-Buller-Drevlak $5\%$ volume-averaged $\beta$ QH configuration (red).}
    \label{fig:coils2}
\end{figure}

We applied the initial, non-planar optimization procedure to the $5\%$ volume-averaged $\beta$ QH configuration of \citet{landremanBullerDrevlak2022}. Five coils per half period were used again for consistency. The currents inside the plasma were accounted for using the virtual casing principle \citep{Shafranov1972, drevlak2005pies}, as implemented numerically by \citet{malhotra2020}.
The weights and target values in the objective function \eqref{eq:obj} were varied to find the values yielding the lowest value for the quantity in \eqref{eq:bn}. The best weight and target values thus found are presented in \autoref{tab:weights2}.

For the resulting coils, the normalized flux surface average of $|\vec{B} \cdot \vec{\hat{n}}|$  (equation \ref{eq:bn}) is $9.7 \times 10^{-4}$.
\autoref{fig:fluxsurfaces2} shows the comparison between the target surface and the surface achieved with the resulting coils, the latter of which was calculated using free-boundary VMEC \citep{vmec1983}. \autoref{fig:countourLBDQH5} shows the magnetic field at the boundary in Boozer coordinates. The coils themselves are shown in \autoref{fig:coils2}, along with the target surface.
The orange curves in \autoref{fig:qsmetrics}(a) and \autoref{fig:qsmetrics}(b) compare quasisymmetry properties of the target configuration and the configuration achieved with the coils, displaying similar levels of relative degradation in quasisymmetry as the Landreman-Paul Precise QH.

With the same fast-particle tracing simulation setup as in \autoref{sec:Landreman-Paulresults}, we obtain alpha-particle loss fractions of $0.0\%$, $2.8\%$ and $10\%$, for particles launched at normalized toroidal flux $s=0.3$, $s=0.5$, $s=0.7$, respectively. For the original (stage I) configuration, the corresponding loss fractions are $0.0\%$, $0.1\%$ and $4\%$.

\begin{table}
\begin{center}
    \begin{tabular}{r c c c c c c}
        & $L$\ & CC Distance & CS Distance & $\kappa$ & MSC &\\
        Weights & 1 & 10 & 10 & 1 & 5 &\\
        Target & 35\,m & 0.9\,m & 1.3\,m & $0.8\,\mathrm{m}^{-1}$ & $0.1\,\mathrm{m}^{-1}$ &\\
        Achieved Value & 35.00\,m & 0.88\,m &	1.29\,m & $0.74\,\mathrm{m}^{-1}$ & $0.11\,\mathrm{m}^{-1}$ &\\
    \end{tabular}
    \caption{Weights placed on terms in the objective function, as well as the target values and the achieved values averaged across the set of coils. For the Landreman-Buller-Drevlak QH configuration described in \autoref{sec:LBDQH5}. The terms in the objective function corresponding to the weights are described in equation \eqref{eq:obj} }
  \label{tab:weights2}
\end{center}
\end{table}

\subsection{Mercier-stable $5\%$ volume-averaged $\beta$ QH\label{sec:mercier}}
\begin{figure}
    \centering
    \includegraphics[width = 1.0\textwidth]{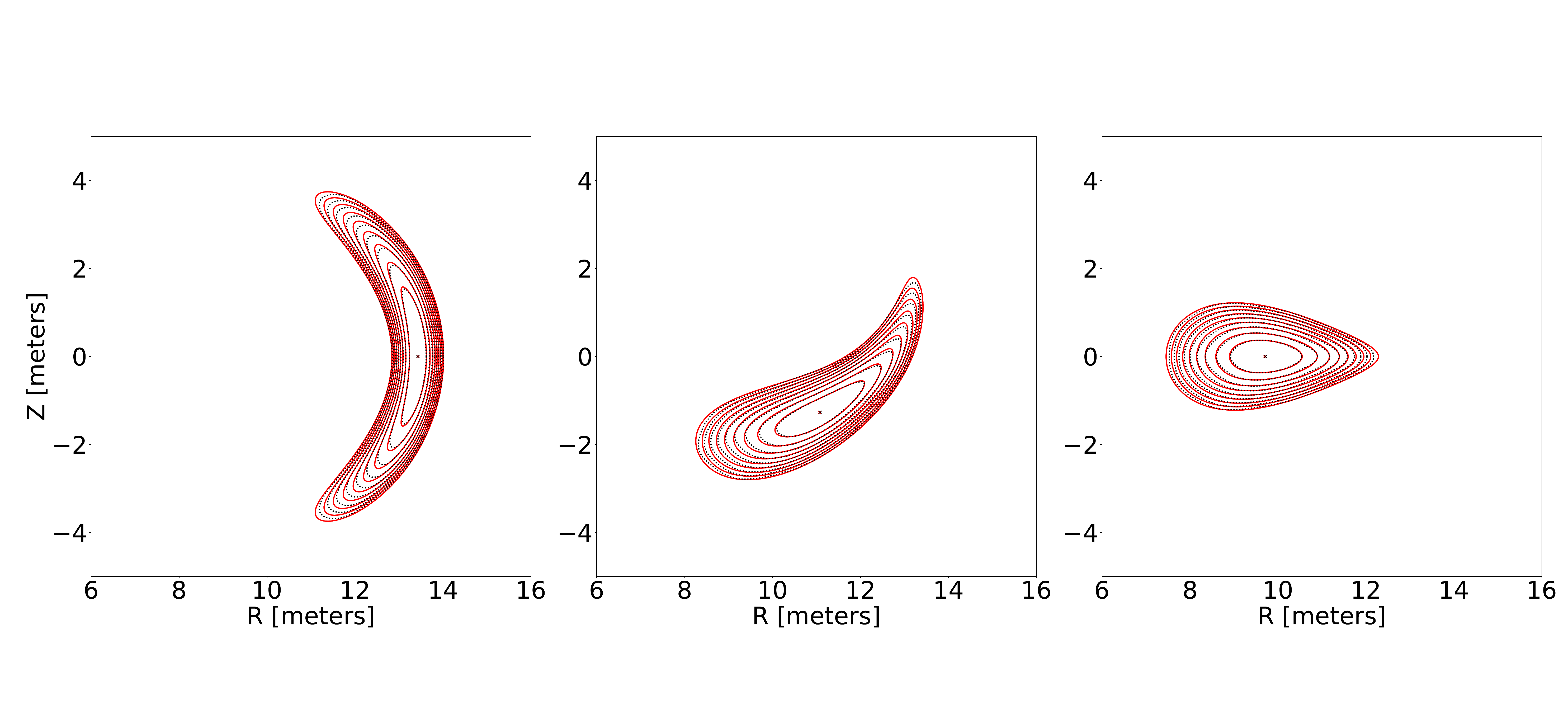}
    \caption{Flux surface comparison of the QH $5\%$ volume-averaged $\beta$ configuration optimized to satisfy the Mercier criterion (red solid), and the surfaces produced by the coils (black dotted).}
    \label{fig:fluxsurfaces3}
\end{figure}

 \begin{figure}
    \centering
    \includegraphics[width=\textwidth]{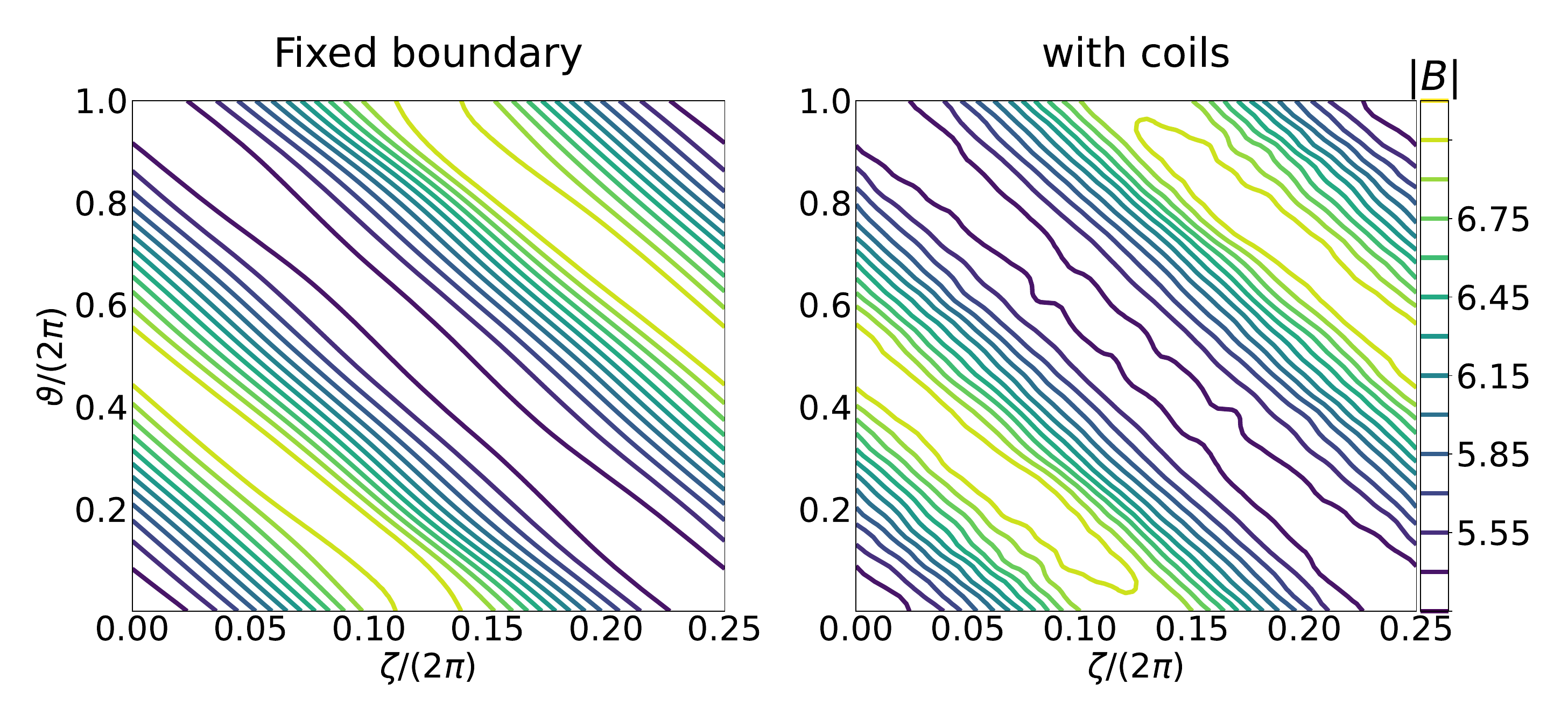}
    \caption{\label{fig:countourMercier} Magnetic field $B$ on the boundary in Boozer coordinates. Left plot:  $5\%$ volume-averaged $\beta$ QH configuration optimized to satisfy the Mercier criterion. Right plot: $B$ produced by the coils presented in \autoref{sec:mercier}.}
\end{figure}

\begin{figure}
    \centering
    \includegraphics[width =0.5\textwidth]{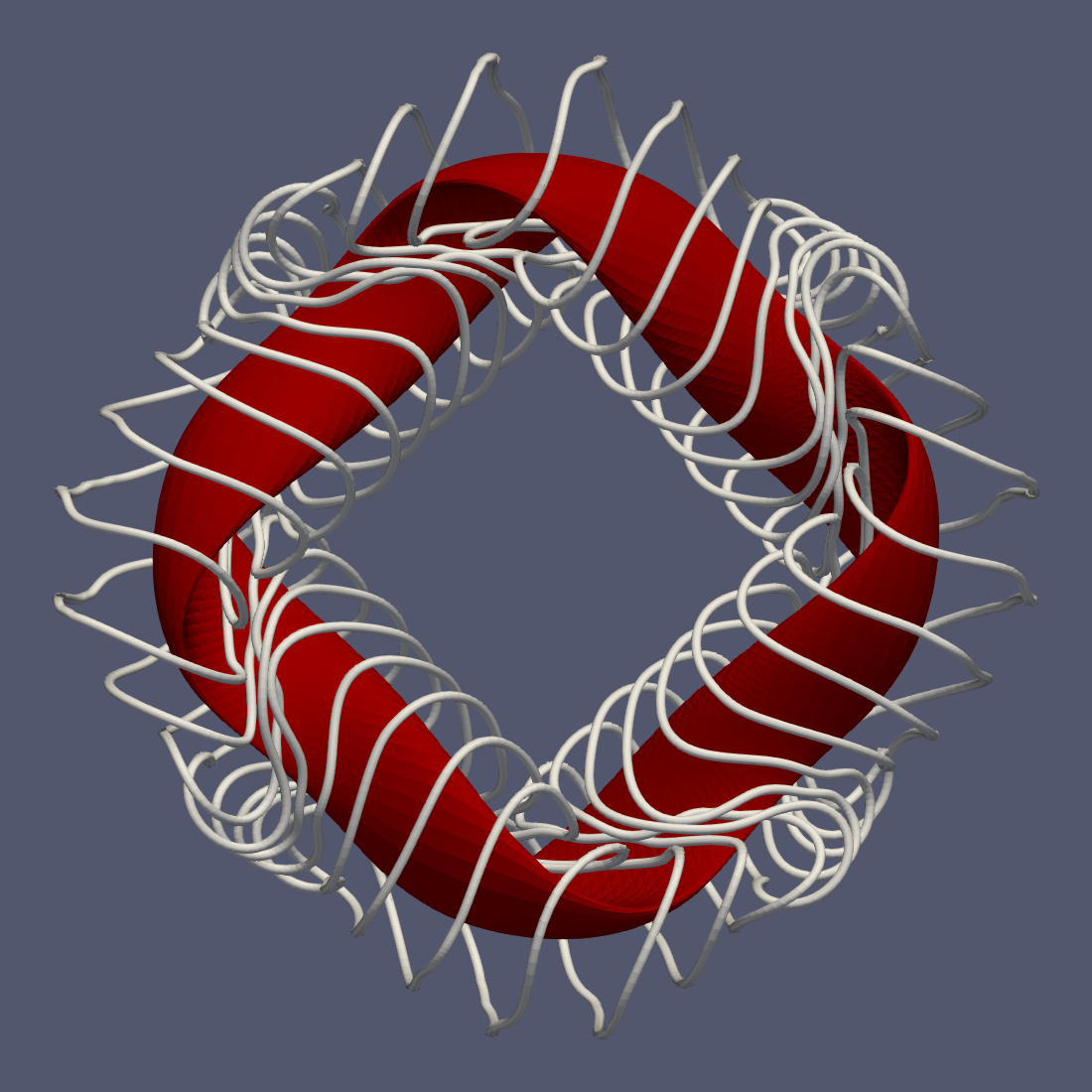}
    \caption{Optimized coil set for the $5\%$ volume-averaged $\beta$ QH Mercier criterion optimized configuration (red).}
    \label{fig:coils3}
\end{figure}

Finally, we optimized coils for a $5\%$ volume-averaged $\beta$ QH configuration that was also optimized to satisfy the Mercier criterion, again using the initial non-planar optimization method with 5 coils per half period.
The weights and target values of the optimization yielding the best coils are presented in \autoref{tab:weights3}. The boundary of this configuration is similar to the
Landreman-Buller-Drevlak QH of the previous section, so the resulting weights are the same, with some of the threshold values slightly tuned. Changing thresholds instead of weights was found to give more predictable results. 
For the resulting coils, the normalized flux surface average of $|\vec{B} \cdot \vec{\hat{n}}|$  (equation \ref{eq:bn}) is $3.9 \times 10^{-3}$. \autoref{fig:fluxsurfaces3} shows the comparison between the target surface and the surface achieved with the coils. \autoref{fig:countourLBDQH5} shows the magnetic field at the boundary in Boozer coordinates. The coils themselves are shown in \autoref{fig:coils3}, along with the target surface.
The green curves in \autoref{fig:qsmetrics}(a) and \autoref{fig:qsmetrics}(b) compare quasisymmetry properties of the target configuration and the configuration achieved with the coils, again displaying similar quasisymmetry degradation as the previous two configurations.

With the same fast-particle tracing simulation setup as in the previous sections, we obtain alpha-particle loss fractions of $3.7\%$, $5.5\%$ and $42\%$, for particles launched at normalized toroidal flux $s=0.3$, $s=0.5$, $s=0.7$, respectively. For particles launched at the outermost radius, about $80\%$ of trapped particles are lost over the $0.2$ seconds simulated, resulting in the large loss fraction. For the stage I configuration, the corresponding loss fractions are $0.4\%$, $3.7\%$, and  $11.5\%$.

\begin{table}
\begin{center}
    \begin{tabular}{r c c c c c c}
        & $L$\,(m) & CC Distance\,(m) & CS Distance\,(m) & $\kappa$\,($\mathrm{m}^{-1}$) & MSC\,($\mathrm{m}^{-1}$) &\\
        Weights (no units) & 1 & 10 & 10 & 1 & 5 &\\
        Target & 35\,m & 1.0\,m & 1.3\,m & $0.8\,\mathrm{m}^{-1}$ & $0.1\,\mathrm{m}^{-1}$ &\\
        Achieved Value & 35.00\,m & 0.94\,m & 1.27\,m & $0.82\,\mathrm{m}^{-1}$ & $0.15\,\mathrm{m}^{-1}$ &\\
    \end{tabular}
    \caption{Weights placed on terms in the objective function as described in equation \eqref{eq:obj} as well as the target values and the achieved values averaged across the set of coils. For the coil set of the $5\%$ volume-averaged $\beta$ QH Mercier criterion optimized configuration described in \autoref{sec:mercier}}
  \label{tab:weights3}
\end{center}
\end{table}

\section{Conclusion} \label{sec:conclusion}
We have found a set of coils for the precise QH configuration in \citet{PhysRevLett.128.035001} that achieves an accurate reconstruction of the magnetic field with $\langle |\vec{B} \cdot \vec{\hat{n}}| \rangle/\langle B \rangle = 0.0006$  while maintaining the targeted coil metrics described in \autoref{sec:tools_and_optimization}. These metrics are compared to NCSX and HSX metrics and shown to be considerably less complex, with this set of coils having half the maximum curvature and a quarter of the mean-squared curvature of NCSX. When perturbing the coils, we have shown that for $\sigma \leq 0.05,\mathrm{m}$, corresponding to coil perturbations of with an average amplitude of $0.13\,\mathrm{m}$,
the fast-particle loss-fraction for particles launched from $s=0.3$ generally stays below $1\%$. The coil set is thus robust to manufacturing error as outlined in \autoref{sec:errortol}. The large allowance in coil perturbation amplitude of about $15\,\mathrm{cm}$ is partly due to the configuration being at reactor scale, and represents a small deviation in comparison to the length of the coils, which is about $35\,\mathrm{m}$.

We have shown that a reasonable number of planar coils alone cannot accurately generate the precise QH magnetic surfaces. The planar coils we found performed significantly worse than the non-planar coils for the given surface, despite having higher curvature and being allowed to have a smaller distances between the coils and surface. It is possible that other configurations may be more easily reproduced with planar coils.

\begin{table}
\begin{center}
    \begin{tabular}{r c c c c c c c c}
         & $L$ & CC d. & CS d.  & $\kappa$ & MSC & eq.\eqref{eq:bn} & $f_{\text{QS}}$ & Losses\\
         & m & m & m  & $\mathrm{m}^{-1}$ & $\mathrm{m}^{-1}$ &  &  & $s=0.5$\\\hline
        LP QH & 35.56 & 1.09 &	1.62 & $0.67$ & $0.07$ &  $6\times10^{-4}$ & 0.00056 (0.000025) & 0.0\% (0.0\%)\\
        LBD QH $5\%$  & 35.00 & 0.88 & 1.29 & $0.74$ & $0.11$ &  $9.7\times10^{-4}$& 0.0090 (0.00051)& 2.8\% (0.1\%)\\
        + Mercier & 35.00 & 0.94 & 1.27 & $0.82$ & $0.15$ &  $3.9\times10^{-3}$& 0.025 (0.0020) & 5.5\% (3.7\%)\\
    \end{tabular}
    \caption{\label{tab:summary} Summary of coil metrics, the accuracy metric \eqref{eq:bn}, two-term quasisymmetry error, and fast-particle loss fraction from the $s=0.5$ surface; for the configurations with coils targeting the Landreman-Paul precise QH, Landreman-Buller-Drevlak $\beta=5\%$ QH, and the $\beta=5\%$ QH optimized to satisfy the Mercier criterion. In parentheses are the corresponding values for the target configurations. See \autoref{tab:weights} for a definitions of the symbols.}
\end{center}
\end{table}

In addition, we optimized coils for two $5\%$ volume-averaged $\beta$ configurations.
\autoref{tab:summary} summarizes the coil metrics, achieved accuracy in reproducing the target boundary, as measured by \eqref{eq:bn}, the two-term quasisymmetry error and the fast-particle loss fractions. The finite-beta configurations have lower coil-coil and coil-surface distance, and larger values of curvature, but still achieve a less accurate reproduction of the plasma boundary when compared to the vacuum configuration. In particular, the coils for the Mercier-optimized configuration have roughly 7 times larger field error (eq \eqref{eq:bn}) than the coils for the vacuum configuration. 
We speculate that this increase can be understood from differences in the gradient scale lengths in the magnetic field of the stage-I target configurations, as measured by the method in \cite{kappel2023prep}.
The magnetic field in the Mercier-stable configuration has the shortest scale length among the three fixed-boundary configurations, while the vacuum configuration has the longest scale length.
Therefore, as discussed in \cite{kappel2023prep}, it is intrinsically most challenging for coils to produce the required field shaping for the Mercier-stable configuration, and least challenging for the vacuum configuration.

Increasing beta to 5\% increases the quasi-symmetry error by about an order of magnitude compared to the vacuum configuration. Further optimizing for the Mercier criterion added another order of magnitude to the error. This increase occurs regardless of whether the configurations are fixed-boundary or reproduced with coils, with the coils adding roughly another order of magnitude in the quasisymmetry error for each configuration. The fast-particle loss fractions are more difficult to interpret, but there is a $2\%$ to $3\%$ absolute increase when comparing the finite-beta configurations produced by coils and their target configurations. The calculated fast-particle losses are still significantly lower than in existing stellarator devices \citep{PhysRevLett.128.035001, Bader_2021}.


Since the optimizations in this work used local rather than global algorithms, better coils may be found given another set of initial conditions, and this could affect the above conclusions.

\paragraph{}
We thank the SIMSOPT team for their support. A.~Wiedman thanks J.~Kappel for assistance with VMEC. This work was supported by the U.S. Department of Energy, Office of Science, Office of Fusion Energy Science, under award number DE-FG02-93ER54197. This research used resources of the National Energy Research Scientific Computing Center (NERSC), a U.S. Department of Energy Office of Science User Facility located at Lawrence Berkeley National Laboratory, operated under Contract No. DE-AC02-05CH11231 using NERSC award FES-ERCAP-mp217-2023. Additional computations were performed on the HPC systems Cobra and Raven at the Max Planck Computing and Data Facility (MPCDF).

\bibliographystyle{jpp}
\bibliography{Placeholder}

\end{document}